\documentclass[aps,prd,superscriptaddress,twocolumn]{revtex4}
\usepackage{bm}
\usepackage{graphicx,amsmath,latexsym,amssymb}
\usepackage{color}
\usepackage{dcolumn}
\begin{document}
\title{Acceleration of a proton along a uniform magnetic field: Casimir momentum of leptons}
\author{M. Donaire} 
\email{manuel.donaire@uva.es, mad37ster@gmail.com}
\affiliation{Departamento de F\'isica Te\'orica, At\'omica y \'Optica and IMUVA,  Universidad de Valladolid, Paseo Bel\'en 7, 47011 Valladolid, Spain}
%\date{28 Jan 2015}

\begin{abstract}
It is commonly assumed that a charged particle does not accelerate linearly along a spatially uniform magnetic field. We show that this is no longer the case if the interaction of the particle with the quantum vacuum is chiral, in which case parity and time-reversal symmetries are simultaneously broken. In particular, this is the situation of an electroweak interacting particle in the presence of a uniform magnetic field. We demonstrate first that, in a spatially uniform and adiabatically time-varying magnetic field, a proton coupled to the leptonic vacuum acquires a kinetic momentum antiparallel to the magnetic field, whereas virtual leptons gain an equivalent Casimir momentum in the opposite direction. Remarkably, leptons remain virtual throughout the process, which means that the proton acceleration is not caused by the recoil associated to the emission of any actual particle. The kinetic energy of the proton is part of its electroweak self-energy, which is provided by the source of magnetic field. %We estimate that, in the magnetic fields of magnetars, protons may reach velocities of the order of $10\:$nm/s. 
 In addition we find that, in a constant and uniform magnetic field, the adiabatic spin-relaxation of a single proton is accompanied by its acceleration along the magnetic field. We estimate that, at the end of the spin-polarization process, the proton reaches a velocity of the order of  $\mu$m/s. The latter finding may lie within the scope of experimental observations. 
\end{abstract}
\maketitle

\section{Introduction}\label{Intro}
 The vacuum state of any quantum field theory contains field fluctuations whose energy is commonly referred to as zero-point energy \cite{Miltonbook,Milonnibook}. Such an energy, which is conjectured to form part of the cosmological constant \cite{Weinberg}, could have observable gravitational effects \cite{Jaffe}. In addition, when quantum field fluctuations couple to material current fluctuations, their interaction energy is finite and can be detected through the  observables of the material degrees of freedom. That is, for instance, through  the forces between conducting plates in the Casimir effect \cite{Casimir}, the van-der-Waals forces between neutral polarizable molecules \cite{vdW}, or the Lamb-shift of neutral atoms \cite{Bethe,Milonnibook}. Beyond QED,  analogous Casimir effects have been addressed in QCD and non-Abelian gauge theories \cite{cavitychromo,Cherenkov}.

On the other hand, homogeneity of space and translation invariance of the vacuum state of any quantum field theory imply that the net momentum of its vacuum field fluctuations is identically zero. However, when quantum field fluctuations couple to material current fluctuations through a Hamiltonian which violates both parity (P) and time-reversal (T), it is symmetry allowed for the field fluctuations to acquire a net virtual momentum.  Conversely, in virtue of total momentum conservation, the material degrees of freedom gain a kinetic momentum of equal strength in the opposite direction, which can be experimentally accessible.

The existence of a virtual momentum of electromagnetic (EM) fluctuations was first shown by Feigel in Ref.\cite{Feigel} --see also Ref.\cite{Croze}. There, P and T symmetries appeared broken by the magneto-optical activity of a chiral medium in a magnetic field. Feigel obtained a net momentum for the fluctuations of the effective EM field, which he referred to as  \emph{Casimir momentum} for its analogy with the energy of the EM fluctuations in the Casimir effect. However, Feigel's semiclassical approach, based on the application of the fluctuation-dissipation theorem to an effective medium, contained unphysical divergences. Later, the QED approach of Ref.\cite{PRLDonaire} revealed that a finite Casimir momentum is indeed carried by the virtual photons which couple to a chiral molecule in the presence of a magnetic field. In contrast to Feigel's, the approach of Ref.\cite{PRLDonaire} was Hamiltonian, with P being broken by the electrostatic potential generated by the chiral distribution of atoms within the molecule, and T being broken by the Zeeman potential of the magnetic field. Further, it was shown that, at a fundamental level, that EM Casimir momentum is due to the differential Doppler-shift experienced by the virtual photons which propagate in the directions parallel and antiparallel to the magnetic field. That differential Doppler shift has its origin in the interaction of the virtual photons with a chromophoric electron within the molecule, which moves at different speeds in the directions parallel and antiparallel to the magnetic field 
as a result of the chiral structure of the molecule \cite{JPCMDonaire}.

The main purpose of this article is to demonstrate that the Casimir momentum is not an effect unique to photons interacting with chiral molecules, but a generic phenomenon in nature. This is so because of the chiral character of the electroweak (EW) interaction of fundamental particles. This implies that not only chiral molecules but also fundamental particles can be accelerated along spatially uniform magnetic fields. In order to demonstrate these facts, we compute the Casimir momentum of the virtual leptons which couple electroweakly to an actual proton in the presence of a magnetic field. %In contrast to EM, parity is not a symmetry of the electroweak (EW) interaction, which is chiral. 
 However, the mechanism through which the proton is accelerated is distinctly different to that which operates upon a chiral molecule. In particular, the aforementioned chirality-induced differential Doppler-shift which operates in the EM case plays no role in the EW case. Quite the contrary, in the EW theory chirality implies an alignment between the spins and the momenta of the species which participate in the EW interaction. As a result, the orientation of the spin of any EW particle induces a net momentum on the actual proton, as well as on the virtual leptons, along the polarization axis. Lastly, spin-polarization is caused by the magnetic field, which breaks T-invariance. 
 
%In its own right, the acceleration of magnetic particles along uniform magnetic fields is in contrast to the generic results of classical electrodynamics. The latter can explain only how charges get accelerated along non-uniform and constant magnetic fields --e.g., in magnetic traps; or the acceleration of charges on the plane perpendicular to a uniform and time-varying magnetic field --e.g., , in a betatron accelerator.
 
Ultimately, we aim at unraveling the transfer of momentum and energy among the proton, the leptonic vacuum and the magnetic field. In this respect, we show that whereas the kinetic energy gained by the proton is a magnetic energy provided by the source of magnetic field, its momentum is provided the virtual leptons of the leptonic vacuum. 
 
 The article is organized as follows. In Sec.\ref{Approach} we describe the fundamentals of our approach as well as the Hamiltonians of the relevant interactions. In Sec.\ref{Adiabatic} we compute the lepton Casimir momentum as well as the velocity reached by a proton along an external magnetic field which is switched on adiabatically. The origin of the proton kinetic energy is clarified. In Sec.\ref{spinrelax}, an analogous calculation is carried out for the case that the proton gets spin-polarized spontaneously while the magnetic field remains constant. The conclusions are summarized in Sec.\ref{Conclusions}, and some relevant comments and remarks are discussed in Sec.\ref{Discussion}.

\section{Fundamentals of the Approach}\label{Approach}

We adopt a quantum field perturbative approach, using the Hamiltonian formalism in Schr\"odinger's representation. Slowly time-varying quantities like momenta and energies are evaluated in the adiabatic limit at zero temperature.
 
Let us denote by $H=H_{0}+W$ the total Hamiltonian of the electroweak system in a spatially uniform magnetic field $\mathbf{B}$, where  $H_{0}=\sum_{\alpha}H^{\alpha}_{D}+H^{\alpha}_{mag}+H_{int}^{N}$ is the nonperturbative Hamiltonian, and $W$ is the perturbative EW interaction. In $H_{0}$, $H^{\alpha}_{D}$ is the Dirac Hamiltonian of free particles of the kind $\alpha$, while  $H^{\alpha}_{mag}$ accounts for their interaction with $\mathbf{B}$. The particles are fundamental leptons, electrons ($e$) and neutrinos ($\nu$); and composite nucleons, protons ($p$) and neutrons ($n$), made of up $(u)$ and down $(d)$ quarks.  Regarding nucleons, the dynamics of their internal and external degrees of freedom are considered decoupled. %\cite{remark}. 
Regarding their external dynamics, protons and neutrons may be considered the two components of an isospin doublet of isospin $I=1/2$, approximately equal masses, and third isospin
components $I_{3}= \pm1/2$, respectively. Hereafter we will label with a script $N$ all those operators which act upon the nucleon doublet.  
 As for the internal dynamics, we treat quarks as constituent quarks with masses $m_{d}\approx m_{u}$. They are bound within the nucleons by some effective strong confining potential which, together with their internal kinetic energy, conform the internal nucleon Hamiltonian $H_{int}^{N}$. Lastly,  $W$ is the EW potential between nucleons and leptons. 
 
%We note that, by considering ...... B CAMBIA EN TIEMPO, POR LO QUE SISTEMA ABIERTO EN EL CASO DE SECCION 3. EN SECCION 4, B FIJO Y PRESENCIA DE WEM.

\subsection{Nonperturbative Hamiltonian of leptons and nucleons}\label{hamil}

As for the Dirac and magnetic Hamiltonians in $H_{0}$, they read in natural units, $\hbar=c=1$,
\begin{align}
H^{\alpha}_{D}&=\int\textrm{d}^{3}R\:\bar{\Psi}_{\alpha}(-i\gamma^{j}\partial_{j}+m_{\alpha})\Psi_{\alpha},\label{HD}\\ 
H^{\alpha}_{mag}&=\int\textrm{d}^{3}R\bar{\Psi}_{\alpha}[-q_{\alpha}\gamma^{j}\tilde{A}_{j}-\frac{\kappa_{\alpha}\mu_{N}}{2}S^{ij}\tilde{F}_{ij}]\Psi_{\alpha},\label{Hmag}
\end{align}
%\begin{equation}
%H^{\alpha}_{D}=\int\textrm{d}^{3}R\:\bar{\Psi}_{\alpha}[-\gamma^{j}(i\partial_{j}+q_{\alpha}A_{j})+m_{\alpha}
%-\frac{\mu^{\alpha}_{A}}{2}S^{ij}F_{ij}]\Psi_{\alpha},\nonumber
%\end{equation}
where $i,j=1,2,3$ are spatial indices, and $\alpha=p,n,e,\nu$. $\Psi_{\alpha}$ are Dirac spinors with charges $q_{\alpha}$ and masses $m_{\alpha}$, and  spatial derivatives for nucleons are with respect to their center of mass, $\mathbf{R}$. In good approximation, we consider that each nucleon interacts with the magnetic field as a whole. In fact, despite the  coupling between the spins of their internal quarks \cite{deRujula,IsgurKarl1978,IsgurKarl1979,Coince,Isgur}, the total spin contribution of these to the magnetic moment of the nucleons is equal to the magnetic moment of a single quark \cite{Karliner}. $\kappa_{\alpha}\mu_{N}/2$ is the anomalous magnetic moment of the nucleons, with $\kappa_{\alpha}/2$ being the anomaly factor and $\mu_{N}$ the nuclear magneton \cite{Scadron,Mao}. $S_{ij}=i[\gamma_{i},\gamma_{j}]/4$ is proportional to the spin operators \cite{Peskin}, $\tilde{F}_{ij}=\partial_{i}\tilde{A}_{j}-\partial_{j}\tilde{A}_{i}$ is the electromagnetic tensor, and our gauge choice for the electromagnetic vector, $\tilde{\mathbf{A}}=-By\hat{\mathbf{x}}$, yields an external magnetic field $\mathbf{B}$ along the $\hat{\mathbf{z}}$-axis, which increases adiabatically from $\mathbf{0}$ to a final value $\mathbf{B}_{0}$. Thus, $H^{\alpha}_{mag}$ evolves adiabatically with $\mathbf{B}$ until $\mathbf{B}=\mathbf{B}_{0}$, and remains constant from them on. Under these conditions, the eigenstates of $H_{0}(\mathbf{B})$ evolve in time with $\mathbf{B}$ towards the \emph{corresponding} eigenstates of the stationary Hamiltonian $H_{0}(\mathbf{B}_{0})$, with $\mathbf{B}$ considered a quasi-stationary parameter in $H_{0}(\mathbf{B})$ at any time. The value of $B$ is such that the inequality $eB\ll m_{d}^{2}$ is assumed throughout this article. 

The eigenstates of $H^{e}_{D}+H^{e}_{mag}(\mathbf{B})$ and $H^{p}_{D}+H^{p}_{mag}(\mathbf{B})$, for electrons and protons respectively, correspond to Landau levels. In the gauge chosen, they are characterized by two quantum numbers, principal number $\mathcal{N}=0,1,2,...$ and spin number $\mathcal{S}=\pm1$; and by the continuous components of their canonical conjugate momenta along $\hat{\mathbf{x}}$ and $\hat{\mathbf{z}}$, $\mathbf{p}$ and $\mathbf{k}$ for protons and electrons, respectively, $|\mathcal{N},\mathcal{S};\mathbf{p}(\mathbf{k})\rangle^{p,e}_{\mathbf{B}}$. 

Virtual electrons and positrons are field fluctuations of the leptonic vacuum, $|\Omega_{l}\rangle_{\mathbf{B}}$, whose energies depend only on $\mathcal{N}$ according to $E^{e}_{\mathcal{N}}=\sqrt{m_{e}^{2}+k_{z}^{2}+2\mathcal{N}eB}$. In terms of creation operators for positrons, $b_{\mathcal{N},\mathcal{S}}^{e\dagger}$, the positron eigenstates read  %and, except for $N=0$, are $two-fold degenerate. 
$|\mathcal{N},\mathcal{S};k_{x},k_{z}\rangle_{\mathbf{B}}^{e}=\sqrt{2E_{\mathcal{N}}^{e}}\:b_{\mathcal{N},\mathcal{S}}^{e\dagger}(k_{x},k_{z})|\Omega_{l}\rangle_{\mathbf{B}}$.

As for the actual proton, its state will be at any time a linear combination of the two lowest Landau levels with $\mathcal{N}=0$ and $\mathcal{S}=\pm1$, $|0,1;\mathbf{p}\rangle^{p}_{\mathbf{B}}$ and $|0,-1;\mathbf{p}\rangle^{p}_{\mathbf{B}}$, with energies  $E^{p}_{\pm}=\sqrt{m_{p}^{2}+p_{z}^{2}+(1\mp1)eB}\mp \kappa_{p}\mu_{N}B/2$, respectively. %$E^{p}_{S}=[(\sqrt{m_{p}^{2}+(1-S)eB}-Sg_{p}\mu_{N}B/2)^{2}+p_{z}^{2}]^{1/2}$ 
In terms of creation operators for protons, $a^{p\dagger}_{\mathcal{N},\mathcal{S}}$, acting upon the hadronic vacuum $|\Omega_{h}\rangle_{\mathbf{B}}$, the eigenstates of the actual proton read $|0,\pm1;\mathbf{p}\rangle^{p}_{\mathbf{B}}=\sqrt{2E_{\pm}^{p}}a^{p\dagger}_{0,\pm1}(p_{x},p_{z})|\Omega_{h}\rangle_{\mathbf{B}}$.

Neutron states of momentum $\mathbf{p}$ experience a Zeeman splitting which depends only on the spin number $\mathcal{S}=\pm1$, $E^{n}_{\pm}=[(\sqrt{m_{n}^{2}+p_{x}^{2}+p_{y}^{2}}\mp \kappa_{n}\mu_{N}B/2)^{2}+p_{z}^{2}]^{1/2}$. Finally, massless neutrinos are free fermions of momentum $\mathbf{q}$ and energy  $E_{q}=q$.  

Expressions of the spinor eigenstates as well as of the fermion fields of all the species are compiled in Appendix \ref{AppA} --cf. \cite{Bhattacharya,Bander,Mao,Broderick}.

\subsection{Internal nucleon effective Hamiltonian}\label{Isgur}

With respect to the internal nucleon Hamiltonian $H_{int}^{N}$, for the sake of simplicity we adopt the simplest effective quantum model to describe the strong confinement of quarks within the nucleons. Such a model consists of considering the three quarks, up and down quarks, as constituent quarks of equal masses, $m_{u}\approx m_{d}$, whose dynamics is nonrelativistic. They are bound together by a strong confining potential which is modelled by a harmonic-oscillator potential, $V_{\textrm{conf}}^{N}$. This is a simplification of the models of Refs.\cite{Coince,IsgurKarl1978,IsgurKarl1979,deRujula}, from which we disregard corrections associated to Coulomb and hyperfine interactions between quarks. 

The relevant EW interaction in our calculations takes place within the nucleon at the location of an active quark $\tilde{\mathbf{r}}$, up for proton, down for neutron, while the remaining quarks rest as spectators [see Fig.\ref{fig1}]. We denote the position vectors of the three constituent quarks of the nucleon by $\mathbf{r}_{u}$,  $\mathbf{r}_{d}$, and $\tilde{\mathbf{r}}$ ,where $\mathbf{r}_{u}$ and $\mathbf{r}_{d}$ refer to the up and down spectator quarks, and $\tilde{\mathbf{r}}$ is the position vector of the active quark in the laboratory frame. Their corresponding conjugate momenta are $\mathbf{p}_{u}$,  $\mathbf{p}_{d}$, and $\tilde{\mathbf{p}}$.  The Jacobi position vectors and corresponding conjugate momenta read \cite{Coince}, 
\begin{align}
\mathbf{R}&=\frac{\mathbf{r}_{u}+\mathbf{r}_{d}+\tilde{\mathbf{r}}}{3},\quad\mathbf{r}_{\rho}=\mathbf{r}_{d}-\mathbf{r}_{u},\quad\mathbf{r}_{\lambda}=\tilde{\mathbf{r}}-\frac{\mathbf{r}_{u}+\mathbf{r}_{d}}{2},\nonumber\\
\mathbf{p}&=\mathbf{p}_{u}+\mathbf{p}_{d}+\tilde{\mathbf{p}},\:\:\mathbf{p}_{\rho}=\frac{\mathbf{p}_{d}-\mathbf{p}_{u}}{2},\:\:\mathbf{p}_{\lambda}=\frac{2\tilde{\mathbf{p}}-\mathbf{p}_{u}-\mathbf{p}_{d}}{3},\nonumber
\end{align}
where $\mathbf{R}$ is the centre of mass of the nucleon and $\mathbf{p}$ its conjugate momentum.  As for the active quark, its position vector  $\tilde{\mathbf{r}}$ reads, in terms of Jacobi's coordinates, $\tilde{\mathbf{r}}=\mathbf{R}+2\mathbf{r}_{\lambda}/3$. 

In the nonrelativistic limit, the total kinetic energy of the nucleon reads, in terms of Jacobi's momenta, 
\begin{align}
\frac{p^{2}_{u}+p_{d}^{2}+\tilde{p}^{2}}{2m_{d}}+\mathcal{O}(eB/m_{d})&=\frac{p^{2}/6+p_{\rho}^{2}+3p_{\lambda}^{2}/4}{m_{d}}\nonumber\\&+\mathcal{O}(eB/m_{d}),\label{kiny}
\end{align}
%T_{N}(\mathbf{p};\mathbf{p}_{\rho};\mathbf{p}_{\lambda})\approx p^{2}/2m_{n}+p_{\rho}^{2}/m_{d}+3p_{\lambda}^{2}/4m_{d}=p^{2}/2m_{n}+T_{N}^{rel}(\mathbf{p}_{\rho};\mathbf{p}_{\lambda}),
%\end{equation}
from which we identify the first term on the right hand side with the kinetic energy of the centre of mass, already accounted for in $H_{D}^{p,n}$; and  the reminder, $p_{\rho}^{2}/m_{d}+3p_{\lambda}^{2}/4m_{d}$, with the internal kinetic energy of the quarks within the internal nucleon Hamiltonian, $H_{int}^{N}$, 
\begin{equation}
H^{N}_{int}= p_{\rho}^{2}/m_{d} + 3p_{\lambda}^{2}/4m_{d} + V^{N}_{\textrm{conf}}(r_{\rho},r_{\lambda}),\label{Hint}
\end{equation}
where $V^{N}_{\textrm{conf}}=m_{d}\omega^{2}(r_{\rho}^{2}/4+r_{\lambda}^{2}/3)-3\omega/2$. The terms of order $eB/m_{d}$ in Eq.(\ref{kiny}) stem from the coupling of the conjugate momenta to $\tilde{\mathbf{A}}$, and are negligible at our order of approximation. Following Refs.\cite{IsgurKarl1978,IsgurKarl1979,Isgur}, approximate values for the constituent quark mass $m_{d}$ and the effective frequency $\omega$ are $m_{d}\approx340$MeV and $\omega\approx250$MeV, respectively. 

Owing to the approximate flavor symmetry of $H_{int}^{N}$ for $m_{d}\approx m_{u}$, $W$ does not affect the internal nucleon dynamics but for the recoil of the active quark.
At zero temperature, the internal nucleon state is the ground state of $H_{int}^{N}$, $|\phi_{int}^{0}\rangle$, whose energy we set to zero. In Jacobian spatial coordinates, its wavefunction is
\begin{equation}
\langle \mathbf{r}_{\lambda},\mathbf{r}_{\rho}|\phi_{int}^{0}\rangle=\frac{\beta^{3}}{3^{3/4}\pi^{3/2}}e^{-\beta^{2}(r_{\lambda}^{2}/3+r_{\rho}^{2}/4)},\quad\beta=\sqrt{\omega m_{d}}.\label{phinot}
\end{equation}
Excited states $|\phi_{int}^{m}\rangle$, $m\geq1$, become rapidly relativistic with energies $E_{int}^{m}=m\omega$. Although $V^{N}_{\textrm{conf}}$ and $|\phi_{int}^{0}\rangle$ are manifestly model-dependent, they set generically an upper limit for the internal momenta of the constituent quarks at $\sim m_{d}$. %\cite{Coince,IsgurKarl1978,IsgurKarl1979,deRujula}.
 
\subsection{The four-fermion effective EW Hamiltonian} 

On top of $H_{0}$, the EW interaction is perturbative. At low energies, it is described by the electroweak V-A potential \cite{FeynmanGellmann,Peskin,betadecay}, 
\begin{align}
&W=\frac{G_{F}}{\sqrt{2}}\int\textrm{d}^{3}R\:J_{h}^{\mu}(\tilde{\mathbf{r}})J^{l}_{\mu}(\tilde{\mathbf{r}})+\textrm{h.c.},\:\:\mu=0,1,2,3,\nonumber\\
&\textrm{where }\:\:J^{l}_{\mu}(\tilde{\mathbf{r}})=\bar{\Psi}_{e}(\tilde{\mathbf{r}})\gamma^{\mu}(\mathbb{I}-\gamma_{5})\Psi_{\nu}(\tilde{\mathbf{r}}),\nonumber\\
&\textrm{and }J_{h}^{\mu}(\tilde{\mathbf{r}})=\bar{\Psi}_{p}(\tilde{\mathbf{r}})\gamma^{\mu}(\mathbb{I}-g_{A}\gamma_{5})\Psi_{n}(\tilde{\mathbf{r}})\nonumber
\end{align}
are the leptonic and hadronic currents respectively, with $g_{A}\simeq1.26$
being the axial-vector coefficient and $G_{F}$ the Fermi constant.  As mentioned earlier, the EW interaction takes place within the nucleon at the location of the active quark, $\tilde{\mathbf{r}}$, while the remaining quarks rest as spectators [Fig.\ref{fig1}].  Important is to notice that, in  low energy phenomena like the one described in this article, the individual hadrons are bound within the nucleons even in the intermediate states of the quantum process. Therefore, no renormalization is needed and there is no need to introduce the fermion-gauge vertex of interaction with W$^{\pm}$ and Z bosons. %The internal energies of the constituent quarks, together with their masses $m_{u}\approx m_{d}$, are accounted for in the nucleon masses $m_{p,n}$ of $H_{D}^{p,n}$\cite{Coince,IsgurKarl1978,IsgurKarl1979,deRujula,SM,remark}, except for the variation of the kinetic energy of the active quark, $\delta T(\tilde{\mathbf{p}})$, associated to its recoil at the EW vertex \cite{SM}.  

Chirality in $W$ manifests in the coupling of left-handed particles to right-handed 
antiparticles. As a result,  there exists a net alignment between the spins and momenta of the hadrons and the leptons which participate in the EW interaction. That explains, for instance, the celebrated experiment of Wu \emph{et al.} \cite{Wu} on the beta decay of 
polarized Co$^{60}$ nuclei, where left-handed electrons were preferably emitted in the direction opposite to the spins of the nuclei 
\cite{betadecay}.

\subsection{Dynamical observables in the adiabatic limit}\label{Dyna}

The fact that the external magnetic field $\mathbf{B}$ which enters Eq.(\ref{Hmag}) increases adiabatically in time, makes it possible to apply the adiabatic theorem to the computation of the expectation value of any operator which evolves adiabatically with $\mathbf{B}$. Let us consider that $\mathbf{B}(\tau)$ increases slowly in time from $\mathbf{0}$ at $\tau=0$ to $\mathbf{B}_{0}$ at some later time $t_{0}>0$. The expression for the expectation value of certain Schr\"odinger operator $\mathcal{O}$ of the system at time $t>0$ reads,
\begin{equation}
\langle\mathcal{O}(t)\rangle=\langle\Phi(0)|\mathbb{U}^{\dagger}(t)\:\mathcal{O}\:\mathbb{U}(t)|\Phi(0)\rangle,\label{equO}
\end{equation}
where $|\Phi(0)\rangle$ is the state of the system at $\tau=0$  and $\mathbb{U}(t)=T\:\exp{[-i\int^{t}_{0}H}\:$d$\tau]$ is the time-propagator in Schr\"odinger's representation. In our perturbative approach, we will expand  $\mathbb{U}(t)$ in powers of $W$. In order to set this explicitly, we rather write 
\begin{equation}
\mathbb{U}(t)=\mathbb{U}_{0}(t)\:T\exp{\left[-i\int_{0}^{t}\textrm{d}\tau\:\mathbb{U}_{0}^{\dagger}(\tau)W\mathbb{U}_{0}(\tau)\right]},\label{U1tot}
\end{equation}
where $\mathbb{U}_{0}(t)=\exp{[-i\int^{t}_{0}H_{0}}\:$d$\tau]$. On the other hand, the adiabatic theorem implies that, if the state of the system starts in an eigenstate of $H$ at $\tau=0$, $\mathbf{B}(0)=\mathbf{0}$, it will evolve towards the corresponding eigenstate of the Hamiltonian $H$ at any latter time $t$ at which $\mathbf{B}(t)\neq\mathbf{0}$ can be considered quasi-stationary. Given the one-to-one correspondence between the value of the field $\mathbf{B}$ and the time $t$, hereafter we will replace  the time-dependence of states and operators on $t$ with their dependence on $\mathbf{B}$. In particular, let us define $|\Phi(0)\rangle=|\Phi\rangle_{\mathbf{0}}\equiv|\Omega_{l}\rangle_{\mathbf{0}}\otimes|\phi_{int}^{N}\rangle\otimes[|0,1;\mathbf{0}\rangle+|0,-1;\mathbf{0}\rangle]^{p}_{\mathbf{0}}/2Lm_{p}^{1/2}$ 
as the initial state of the system, where we use the nomenclature of Sec.\ref{hamil} to denote the state of  an unpolarized proton at rest and the leptonic vacuum, at zero magnetic field, with $L\rightarrow\infty$ being a normalization length. The adiabatic theorem implies that, at a later time $t$, this state will evolve towards its corresponding eigenstates of $H(\mathbf{B})$ with a quasi-stationary magnetic field $\mathbf{B}$. Using the nomenclature in Sec.\ref{hamil},
\begin{align}
&|\Phi\rangle_{\mathbf{B}}=\mathbb{U}(\mathbf{B})|\Phi\rangle_{\mathbf{0}}\label{PB}\\
&\simeq |\Omega_{l}\rangle_{\mathbf{B}}\otimes|\phi_{int}^{N}\rangle\otimes\bigl[\bigl|0,1;\langle\mathbf{P}_{N}\rangle\bigl\rangle+|0,-1;\langle\mathbf{P}_{N}\rangle\bigl\rangle\bigr]^{p}_{\mathbf{B}}/2L\sqrt{E_{0}^{p}}.\nonumber
\end{align}
 Note that, in order to keep the computation of $\langle\mathcal{O}(t)\rangle$ at lowest order in $W$, $|\Phi\rangle_{\mathbf{B}}$ has been approximated to $\mathbb{U}_{0}(\mathbf{B})|\Phi\rangle_{\mathbf{0}}$ but for  the shift $\langle\mathbf{P}_{N}\rangle_{\mathbf{B}}$ in the linear momentum that the nucleon ($N$) acquires as a result of the action of $W$ in Eq.(\ref{U1tot}). This way, in our approximation,  $|\Phi\rangle_{\mathbf{B}}$ remains being an eigenstate of $H_{0}(\mathbf{B})$. 

Next, the expectation value $\langle\mathcal{O}(t_{0})\rangle=\langle\mathcal{O}\rangle_{\mathbf{B}_{0}}$ can be computed out of the integration of the variations $\delta\langle\mathcal{O}\rangle_{\mathbf{B}}$ with respect to the adiabatic variations of the magnetic field, $\delta\mathbf{B}$, from $\mathbf{B}=\mathbf{0}$ to $\mathbf{B}_{0}$. In order to compute those variations it is convenient to define a new time-evolution operator which takes account of the adiabatic evolution of the system upon which the operator $\mathcal{O}$ applies. At leading order in $W$ we define
\begin{equation}
\tilde{\mathbb{U}}_{\mathbf{B}}(t)\equiv\tilde{\mathbb{U}}_{\mathbf{B}}^{0}(t)-i\int_{-\infty}^{t}\textrm{d}\tau\:\tilde{\mathbb{U}}_{\mathbf{B}}^{0}(t-\tau)W\tilde{\mathbb{U}}_{\mathbf{B}}^{0}(\tau)e^{\eta\tau},\:\eta\rightarrow0^{+},\label{eta}
\end{equation}
where $\tilde{\mathbb{U}}_{\mathbf{B}}^{0}(\tau)=e^{-iH_{0}(\mathbf{B})\tau}$, for $\mathbf{B}$ stationary. Using this definition we can write the adiabatic variation $\delta\langle\mathcal{O}\rangle$ with respect to an adiabatic variation $\delta\mathbf{B}$, at certain value of the magnetic field $\mathbf{B}$, as 
\begin{align}
\delta\langle\mathcal{O}\rangle_{\mathbf{B}}&=\Bigl[\:_{\mathbf{0}}\langle\Phi|\frac{\delta}{\delta\mathbf{B}}\Bigl(\mathbb{U}^{\dagger}(\mathbf{B})\tilde{\mathbb{U}}^{\dagger}_{\mathbf{B}}(t)\Bigr)\:\mathcal{O}\:\tilde{\mathbb{U}}_{\mathbf{B}}(t)\mathbb{U}(\mathbf{B})|\Phi\rangle_{\mathbf{0}}\nonumber\\
&+\:_{\mathbf{0}}\langle\Phi|\mathbb{U}^{\dagger}(\mathbf{B})\tilde{\mathbb{U}}^{\dagger}_{\mathbf{B}}(t)\:\mathcal{O}\:\frac{\delta}{\delta\mathbf{B}}\Bigl(\tilde{\mathbb{U}}_{\mathbf{B}}(t)\mathbb{U}(\mathbf{B})\Bigr)|\Phi\rangle_{\mathbf{0}}\Bigr]\delta\mathbf{B}.\label{dO}
\end{align}
Therefore, in the adiabatic limit, at time $t_{0}$, Eq.(\ref{equO}) reads 
\begin{equation}
\langle\mathcal{O}(t_{0})\rangle=\int_{\mathbf{0}}^{\mathbf{B}_{0}}\delta\langle\mathcal{O}\rangle_{\mathbf{B}}.\label{Otot}
\end{equation}
It is worth noting that the formalism presented here is a generalisation of the so-called Hellmann-Feynman-Pauli theorem which applies to the case that $\mathcal{O}=H_{0}(\mathbf{B})$. That is, to the computation of the energy of a system whose Hamiltonian depends on an adiabatic parameter \cite{Pines}. 

We finalize this section with some comments relevant to the subsequent calculations. In the first place, we note that since $|\Phi\rangle_{\mathbf{B}}$ is a stationary state throughout the adiabatic variation of $\mathbf{B}$,  Eqs.(\ref{eta}) and (\ref{dO}) imply that $\delta\langle\mathcal{O}\rangle_{\mathbf{B}}$ can be computed using stationary perturbation theory upon the time-independent Hamiltonian $H_{0}(\mathbf{B})+W$. Further, provided that the operator $\mathcal{O}$ is independent of the adiabatic parameter $\mathbf{B}$, $\langle\mathcal{O}(t_{0})\rangle$ can be computed using stationary perturbation theory upon the time-independent Hamiltonian $H_{0}(\mathbf{B}_{0})+W$. That is, if $\delta\mathcal{O}/\delta\mathbf{B}=\mathbf{0}$, it holds that, for $t\gtrsim t_{0}$, 
\begin{equation}
\langle\mathcal{O}(t)\rangle=\:_{\mathbf{B}_{0}}\langle\Phi|\:\tilde{\mathbb{U}}^{\dagger}_{\mathbf{B_{0}}}(t)\:\mathcal{O}\:\tilde{\mathbb{U}}_{\mathbf{B}_{0}}(t)\:|\Phi\rangle_{\mathbf{B}_{0}},\quad t\gtrsim t_{0}.\label{Ototstat}
\end{equation}
In fact, from Eq.(\ref{eta}) it is plain that $\langle\mathcal{O}(t)\rangle$ in Eq.(\ref{Ototstat}) does not depend explicitly on $t$. 

%Lastly, if the computation of $\langle\mathcal{O}(t_{0})\rangle$ is performed perturbatively up to certain order in $W$, the functional derivatives with respect to $\mathbf{B}$ in Eq.(\ref{dO}) must take into account the order in $W$ of the factor $\delta\mathbb{U}^{\dagger}(\mathbf{B})$. We will see that this is of particular relevance to incorporation of $\langle\mathbf{P}_{N}\rangle$ in Eq.(\ref{PB}).

\section{Lepton Casimir momentum  in an adiabatically time-varying magnetic field}\label{Adiabatic}

%CALCULAR LA FUERZA COMO DERIVADA TEMPORAL DEL MOMENTO Y POR TANTO LA FUERZA RESPECTO AL LAMB-SHIFT DE EW 
In this Section we show by direct computation that the application of an adiabatically increasing magnetic field $\mathbf{B}$ on an initially 
unpolarized proton at rest, makes virtual leptons acquire a net linear momentum parallel to $\mathbf{B}$ while the proton accelerates in the opposite direction. In addition, we demonstrate that the kinetic energy gained by the proton is part of the magnetic energy provided by the source of the magnetic field. 

\subsection{Lepton Casimir momentum}\label{Adiamomentum}

\begin{figure}[h]
\includegraphics[height=5.5cm,width=8.6cm,clip]{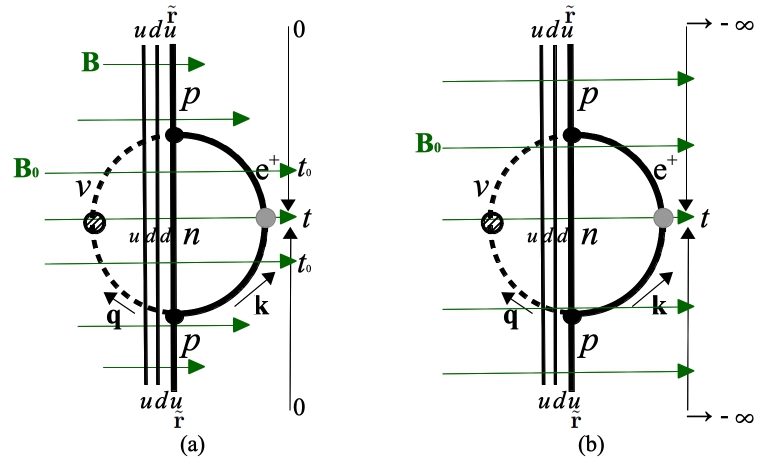}
\caption{(a) Diagrammatic representation of $\langle(\mathbf{K}_{e^{+}}+\mathbf{Q}_{\nu})(t)\rangle$ at order $W^{2}$ considering the time evolution of $\mathbf{B}$.  Time runs along the vertical. The magnetic field increases from $\mathbf{0}$ at time $0$ to $\mathbf{B}_{0}$ at $t_{0}$. Lepton kinetic momentum operators apply at the observation time $t\gtrsim t_{0}$. Vertices stand for $W$; dashed and grey circles at time $t$ depict the kinetic momentum operators of neutrinos and positrons, respectively. (b) Diagrammatic representation of $\langle(\mathbf{K}_{e^{+}}+\mathbf{Q}_{\nu})(t)\rangle$  according to Eqs.(\ref{equ}) and (\ref{equationmother}) using the adiabatic approximation. The value of the magnetic field remains constant at $\mathbf{B}_{0}$, and the initial state of the system is $|\Phi\rangle_{\mathbf{B}_{0}}$ at $\tau\rightarrow-\infty$.}\label{fig1} %(b)\&(c) Schematic representation of the P and T symmetry-breaking processes in the transfer of momentum between the virtual leptons and the nucleon according to Eqs.(\ref{f1a})\&(\ref{f2a}), respectively. (d) Schematic representation of the nucleon-leptonic vacuum system considered as a whole, with null total momentum along $\mathbf{B}_{0}$.
\end{figure}

In order to compute the Casimir momentum we apply the formalism of Sec.\ref{Dyna} to the expectation value of the leptonic momentum as the magnetic field increases from $\mathbf{0}$ to its final value $\mathbf{B}_{0}$, at leading order in $\mathbf{B}_{0}$. That is, the quantity of our interest is $\langle\mathbf{K}_{e^{+}}+\mathbf{Q}_{\nu}\rangle_{\mathbf{B}_{0}}$, where $\mathbf{K}_{e}$ and $\mathbf{Q}_{\nu}$ are the kinetic momentum operators of positrons and neutrinos, respectively,
\begin{align}
\mathbf{Q}_{\nu}&=\int\textrm{d}^{3}r\:\Psi^{\dagger}_{\nu}(\mathbf{r})(-i\mathbf{\nabla})\Psi_{\nu}(\mathbf{r})|_{\textrm{neutrinos}},\nonumber\\
\mathbf{K}_{e^{+}}&=\int\textrm{d}^{3}r\:[\Psi^{\dagger}_{e}(\mathbf{r})(-i\mathbf{\nabla})\Psi_{e}(\mathbf{r})+e\tilde{\mathbf{A}}]|_{\textrm{positrons}}.
\end{align}
The expressions for $\mathbf{K}_{e^{+}}$ and $\mathbf{Q}_{\nu}$ in terms of creation and annihilation operators of positrons, $b_{\mathcal{N},\mathcal{S}}^{e(\dagger)}(k_{x},k_{z})$, and neutrinos, $a^{\nu(\dagger)}_{\mathcal{S}}(\mathbf{q})$, are derived in Appendix \ref{KQ}. Since the operators $\mathbf{K}_{e^{+}}$ and $\mathbf{Q}_{\nu}$ do not depend on $\mathbf{B}$, from Eq.(\ref{Ototstat}), $\langle(\mathbf{K}_{e^{+}}+\mathbf{Q}_{\nu})(t)\rangle$ reads, for $t\gtrsim t_{0}$,
\begin{equation}
\langle(\mathbf{K}_{e^{+}}+\mathbf{Q}_{\nu})(t)\rangle=\:_{\mathbf{B}_{0}}\langle\Phi|\:\tilde{\mathbb{U}}^{\dagger}_{\mathbf{B_{0}}}(t)(\mathbf{K}_{e^{+}}+\mathbf{Q}_{\nu})\tilde{\mathbb{U}}_{\mathbf{B}_{0}}(t)\:|\Phi\rangle_{\mathbf{B}_{0}},\label{equ}
\end{equation}
where $|\Phi\rangle_{\mathbf{B}_{0}}$ is given in Eq.(\ref{PB}). We compute Eq.(\ref{equ}) at leading order in $W$, i.e., $\mathcal{O}(W^{2})$, using Eq.(\ref{eta}).  Its diagrammatic representation 
 is given in Fig.\ref{fig1}(b). There, an actual proton %, stable for $\mu_{N}B\ll m_{n}-m_{p}$, 
emits a virtual positron-neutrino pair in the presence of an external magnetic field $\mathbf{B}_{0}$. The momentum operators of leptons apply at the observation time $t\gtrsim t_{0}$. 

The chiral nature of $W$ along with the magnetic field $\mathbf{B}$ break the mirror symmetry with respect to the $xy$-plane. Therefore, the only nonvanishing component of  $\langle(\mathbf{K}_{e^{+}}+\mathbf{Q}_{\nu})(t)\rangle$ can be that along $\mathbf{B}_{0}\parallel\hat{\mathbf{z}}$. In addition, translation invariance of $H$ along the magnetic field implies conservation of total momentum in that direction, i.e., $\hat{\mathbf{z}}\cdot\langle\mathbf{P}_{N}(t)\rangle=-\hat{\mathbf{z}}\cdot\langle(\mathbf{K}_{e^{+}}+\mathbf{Q}_{\nu})(t)\rangle$. Therefore, if the leptonic momentum presents a non-zero component along $\mathbf{B}$, we can be certain that the  proton accelerates in the opposite direction. $\langle\mathbf{P}_{N}(t)\rangle$ is the quantity that can be accessed experimentally, whereas $\langle(\mathbf{K}_{e^{+}}+\mathbf{Q}_{\nu})(t)\rangle$ is what can be computed.

%We consider here that the observation time $t_{0}$ is not yet long enough for the spin-relaxation of the proton. 
 Applying these considerations to Eq.(\ref{equ}), the Casimir momentum of leptons at time $t\gtrsim t_{0}$ is, at $\mathcal{O}(W^{2})$, %\cite{Peskin},
\begin{align}
&\langle(\mathbf{K}_{e^{+}}+\mathbf{Q}_{\nu})(t)\rangle=\frac{G_{F}^{2}}{2}\hat{\mathbf{z}}\int\textrm{d}^{3}R\textrm{d}^{3}R'\:_{\mathbf{B}_{0}}\langle\Phi|J_{h}^{\mu}(\tilde{\mathbf{r}})J^{l}_{\mu}(\tilde{\mathbf{r}})\nonumber\\&\times\frac{(\mathbf{K}_{e^{+}}+\mathbf{Q}_{\nu})\cdot\hat{\mathbf{z}}}{[H_{0}(\mathbf{B}_{0})-E_{0}^{p}]^{2}}J^{\rho\dagger}_{\mu}(\tilde{\mathbf{r}}')J_{h}^{\rho\dagger}(\tilde{\mathbf{r}}')|\Phi\rangle_{\mathbf{B}_{0}},\quad t\gtrsim t_{0},\label{equationmother}
%&|\Phi(t_{0})\rangle=T\:\exp{\Bigl[-i\int^{t_{0}}_{0}H_{0}\:d\tau\Bigr]}|\Phi(0)\rangle\nonumber\\
%&\qquad\:\:=|\Omega_{l}\rangle_{\mathbf{B}_{0}}\otimes|\phi_{int}^{N}\rangle\otimes\bigl[|0,1\rangle+|0,-1\rangle\bigr]^{p}_{\mathbf{B}_{0}}/2L\sqrt{E_{0}^{p}},\nonumber
\end{align}
where $E_{0}^{p}=m_{p}+\mu_{N}B_{0}$ and 
\begin{align}
\mathbf{Q}_{\nu}\cdot\hat{\mathbf{z}}&=\int\frac{\textrm{d}^{3}q}{(2\pi)^{3}}\:q_{z}\sum_{\mathcal{S}=\pm1}a^{\nu\dagger}_{\mathcal{S}}(\mathbf{q})a^{\nu}_{\mathcal{S}}(\mathbf{q}),\nonumber\\
\mathbf{K}_{e^{+}}\cdot\hat{\mathbf{z}}&=\int\frac{\textrm{d}k_{x}\textrm{d}k_{z}}{(2\pi)^{2}}k_{z}\sum_{\mathcal{S}=\pm1}\sum_{\mathcal{N}=0}b^{e\dagger}_{\mathcal{N},\mathcal{S}}(k_{x},k_{z})b^{e}_{\mathcal{N},\mathcal{S}}(k_{x},k_{z}).\nonumber
\end{align}
The leptonic and hadronic currents flanking the energy denominator of Eq.(\ref{equationmother}) give rise to quadratic vacuum fluctuations for the case of leptons, and quadratic hadron fluctuations upon the external proton state $\bigl[\bigl|0,1;\langle\mathbf{P}_{N}\rangle\bigl\rangle+|0,-1;\langle\mathbf{P}_{N}\rangle\bigl\rangle\bigr]^{p}_{\mathbf{B}_{0}}/2L\sqrt{E_{0}^{p}}$. Explicit expressions of these fluctuations are written down in Appendix \ref{AppC}. It is of note that $\langle(\mathbf{K}_{e^{+}}+\mathbf{Q}_{\nu})(t)\rangle$ remains constant for $t\gtrsim t_{0}$ as long as the state $|\Phi\rangle_{\mathbf{B}_{0}}$ remains stationary too. 
%and $\sum_{\alpha}H_{mag}^{\alpha}$ within $H_{0}$ in the denominator of Eq.(\ref{equationmother}) is evaluated at $\mathbf{B}_{0}$.  

In Eqs.(\ref{equ}) and (\ref{equationmother}), $\sum_{\alpha}H_{mag}^{\alpha}$ in $H_{0}(\mathbf{B}_{0})$ causes an  effective orientation of the spin of each magnetic particle along $\mathbf{B}_{0}$, $\alpha=e,n,p$. On the other hand, the chiral potential $W$ induces an alignment between the spins of the particles and their momenta. Both effects together result in a net alignment of the linear momenta of leptons and hadrons along $\mathbf{B}_{0}$ and, in turn, in a nonvanishing value of $\langle(\mathbf{K}_{e^{+}}+\mathbf{Q}_{\nu})(t)\rangle$.  Its final expression reads 
\begin{widetext}
\begin{align} 
&\langle(\mathbf{K}_{e^{+}}+\mathbf{Q}_{\nu})(t)\rangle\simeq\textrm{Re}\int\textrm{d}y\:\textrm{d}y'\int\frac{\textrm{d}p_{y}}{2\pi}\int\frac{\textrm{d}^{3}q}{(2\pi)^{3}}\int\frac{\textrm{d}k_{x}\textrm{d}k_{z}}{(2\pi)^{2}}e^{i(p_{y}+q_{y})(y-y')}(k_{z}+q_{z})\hat{\mathbf{z}}\sum_{\mathcal{N},m=0}^{\infty}\frac{G_{F}^{2}}{q\:\mathcal{E}_{\mathcal{N},m}^{2}E^{e}_{\mathcal{N}}}\Bigl|\langle\tilde{\phi}_{int}^{0}|\tilde{\phi}_{int}^{m}\rangle\Bigr|^{2}\nonumber\\
&\times\Bigg\{I_{0}(\xi_{p})I_{0}(\xi^{'}_{p})(\mathbf{k}\cdot\langle\delta\mathbf{S}_{e^{+}}^{\mathcal{N}}\rangle)\left[q\:(1+3g^{2}_{A})+\frac{\mathbf{q}\cdot\mathbf{p}}{m_{n}}(g_{A}-1)^{2}\right]
+I_{\mathcal{N}}(\xi_{e^{+}})I_{\mathcal{N}}(\xi^{'}_{e^{+}})\bigg[ (\mathbf{k}\cdot\langle\delta\mathbf{S}_{p}\rangle)q\:(g_{A}-1)^{2}\label{f1a}\\
&+ (\mathbf{q}\cdot\langle\delta\mathbf{S}_{p}\rangle) E^{e}_{\mathcal{N}}(g_{A}+1)^{2}\bigg]+I_{\mathcal{N}}(\xi_{e^{+}})I_{\mathcal{N}}(\xi^{'}_{e^{+}})I_{0}(\xi_{p})I_{0}(\xi^{'}_{p})\bigg[(\mathbf{k}\cdot\langle\delta\mathbf{S}_{n}\rangle)2q\:(1+3g^{2}_{A}+4g_{A})+ (\mathbf{q}\cdot\langle\delta\mathbf{S}_{n}\rangle)E^{e}_{\mathcal{N}}(g^{2}_{A}-1)\bigg]\Bigg\},\nonumber\\
&\textrm{where}\:\:\langle\delta\mathbf{S}_{e^{+}}^{\mathcal{N}}\rangle=\frac{e\mathbf{B}_{0}}{2(E^{e}_{\mathcal{N}})^{2}}I_{\mathcal{N}}(\xi_{e^{+}})I_{\mathcal{N}}(\xi^{'}_{e^{+}}),\:\:\langle\delta \mathbf{S}_{p}\rangle=\frac{\mu_{N}\mathbf{B}_{0}}{2m_{p}}[I_{0}(\xi_{p})I_{0}(\xi^{'}_{p})+I_{1}(\xi_{p})I_{1}(\xi^{'}_{p})],\textrm{ and }\langle\delta \mathbf{S}_{n}\rangle=\frac{|\kappa_{n}|\mu_{N}\mathbf{B}_{0}}{4m_{n}}\label{spins}
\end{align}
\end{widetext}
are, respectively, the net spin density of any pair of consecutive positron Landau levels, $|\mathcal{N},-1;\mathbf{k}\rangle^{e}_{\mathbf{B}_{0}}$ and $|\mathcal{N}+1,1;\mathbf{k}\rangle^{e}_{\mathbf{B}_{0}}$; the net spin density of the two lowest proton Landau levels, $|0,1;\mathbf{0}\rangle^{p}_{\mathbf{B}_{0}}$ and $|0,-1;\mathbf{0}\rangle^{p}_{\mathbf{B}_{0}}$; and the net spin of any pair of neutron states with equal momenta and different spin numbers, -1 and +1. Their equations are written in Appendix \ref{spines}. In Eq.(\ref{f1a}), $y$ and $y'$ are the coordinates of the centre of mass of the nucleon along $\hat{\mathbf{y}}$; $\mathbf{k}$ is the canonical momentum of the virtual positrons; $\mathbf{q}$ is the momentum of the virtual neutrinos; and $\mathbf{p}$ is the momentum of the intermediate neutron. At each vertex, total momentum fails to be conserved on the plane perpendicular to $\mathbf{B}_{0}$. With our choice of gauge, conservation holds however along the axis $\hat{\mathbf{x}}$ and $\hat{\mathbf{z}}$, $p_{x,z}=-(k_{x,z}+q_{x,z})$. Concerning the energy denominators, 
for $eB_{0}\ll m^{2}_{d}$, $\mathcal{E}_{\mathcal{N},m}\approx(k_{z}^{2}+q^{2}+2\mathcal{N}eB_{0})^{2}/2m_{d}+q_{z}k_{z}/m_{d}+q+E_{\mathcal{N}}^{e}+m_{n}-m_{p}+m\omega$ is the energy difference between the initial and intermediate states. The derivation of the kinetic terms is detailed in Appendix \ref{AppD1}. The functions $I_{\mathcal{N}}(\xi)=(eB_{0})^{1/4}H_{\mathcal{N}}(\xi)e^{-\xi^{2}/2}/\pi^{1/4}\sqrt{2^{\mathcal{N}}\mathcal{N}!}$ are the charge distribution functions of the Landau levels, with $H_{\mathcal{N}}$ the Hermite polynomial of order $\mathcal{N}$, $\xi^{(')}_{p}=\sqrt{eB_{0}}y^{(')}$ for protons, and $\xi^{(')}_{e^{+}}=\sqrt{eB_{0}}y^{(')}+k_{x}/\sqrt{eB_{0}}$ for positrons. Finally, $|\langle\tilde{\phi}_{int}^{0}|\tilde{\phi}_{int}^{m}\rangle|^{2}$ is the square of the form factor between the internal nucleon ground state and the $m^{th}$ state,
\begin{equation}
\langle\tilde{\phi}_{int}^{0}|\tilde{\phi}_{int}^{m}\rangle=\langle\phi^{0}_{int}|e^{\pm2i[\mathbf{r}_{\lambda}\cdot(\mathbf{k}+\mathbf{q})+r_{\lambda}^{y}(\pm\sqrt{2\mathcal{N}eB}-k_{y})]/3}|\phi_{int}^{m}\rangle.\nonumber
\end{equation}
For the nonrelativistic harmonic oscillator of Eq.(\ref{Hint}), the form factors posses a Gaussian profile in momentum space  which, in an effective manner, sets a cutoff for the momentum integrals at $\sim m_{d}$ --see Eq.(\ref{formy}). In particular, for $m=0$, 
\begin{equation}
\langle\tilde{\phi}_{int}^{0}|\tilde{\phi}_{int}^{0}\rangle\simeq e^{-[(k_{x}+q_{x})^{2}+(k_{z}+q_{z})^{2}+q_{y}^{2}+2\mathcal{N}eB]/6\beta^{2}}.%e^{\frac{\mp q_{y}\sqrt{2\mathcal{N}eB}}{3\alpha^{2}}},
\nonumber
\end{equation}
 
As anticipated, Eq.(\ref{spins}) shows that the effective spin-polarization of protons, positrons and neutrons is proportional to the magnetic field. In addition, the alignment of the momenta of virtual positrons and neutrinos with the spins reflects in the pseudo-scalar products of spins and momenta in the integrand of Eq.(\ref{f1a}). As in Wu's experiment \cite{Wu,betadecay}, that causes virtual positrons and neutrinos to be emitted and absorbed with different probabilities in the directions parallel and antiparallel to $\mathbf{B}_{0}$. In our case, in virtue of total momentum conservation along $\mathbf{B}_{0}$, that results in a net transfer of momentum to the actual proton. %The dominant terms are those proportional to $\langle\delta\mathbf{S}_{e^{+}}^{\mathcal{N}}\rangle$. 
Passing the sums to continuum integrals and using the nonrelativistic harmonic-oscillator model of Isgur \& Karl for the internal eigenstates of the nucleon \cite{Coince,IsgurKarl1978,IsgurKarl1979}, we arrive at %XXXXX HABLAR AQUI DE COMO SE HACE LA SUMA SOBRE TODOS LOS ESTADO LIGADOS DEL NUCLEON Y COMO EL QUE MAS IMPORTA ES EL GROUND STATE
\begin{align} 
&\langle(\mathbf{K}_{e^{+}}+\mathbf{Q}_{\nu})(t)\rangle_{\mathbf{B}_{0}}\approx\frac{G_{F}^{2}m_{d}^{3}}{160\pi^{4}}\Big[(1+3g_{A}^{2})-4\frac{m_{d}}{m_{n}}(g_{A}-1)^{2}\nonumber\\
&\qquad\qquad\quad\:\:+2\frac{m^{2}_{d}}{m^{2}_{n}}(10g^{2}_{A}+3g_{A}+9)\Big]e\mathbf{B}_{0},\:t\gtrsim t_{0},\label{f1b}\\
&\langle\mathbf{v}_{p}\rangle_{\mathbf{B}_{0}}\approx-1.5\cdot10^{-15}\frac{e\mathbf{B}_{0}}{m_{d}m_{p}},\textrm{ in natural units }c=1,\nonumber
\end{align}
for the leptonic Casimir momentum and the proton velocity, respectively. It is of note that, as for the case of the calculation of the Lamb-shift for the hydrogen atom, it is the inclusion of retardation factors in the lepton fields, together with the internal wavefunction of the nucleon, which makes the result finite --cf. Refs.\cite{Milonnibook,Bethe} and Appendix \ref{AppD}. In addition, the nonrelativistic approximation implies the restriction of the sum over intermediate internal nucleon states to $m=0$ in Eq.(\ref{f1a}). It is explained inf Appendix \ref{AppD2} that the contribution of excited states is negligible at our order of approximation. 

From the result of Eq.(\ref{f1b}) we find that the velocity that a proton can reach under the action of the strongest magnetic fields in laboratory cannot exceed $10^{-20}$m/s, which is undetectable in experiments. However, for the typical values of $B_{0}$ within magnetars, between $10^{11}$T and $10^{14}$T \cite{Khalilov2005}, we obtain $10^{-2}$nm/s$<v_{p}<10\:$nm/s, which might have an impact on the distribution of matter in those extremely dense objects \cite{Mao}. 

\subsection{Nature of the proton kinetic energy}\label{AdiaEnergy}
%CREO MEJOR DEJAR SOLO ANUNCIADO QUE LA ENERGIA VIENE DE LA FUENTE DE CAMPO MAGNETICO Y QUE, ANALOGAMENTE AL CASO DE LA ENERGIA DE LA MOLECULA CHIRAL, ES PARTE DE LA EW SELF-ENERGY. FALTA PROBAR EL ORIGEN DE LA ENERGIA DE POLARIZACION. QUE DICHA ENERGIA DEBE SER MAGNETICA, VIENE DEL TEOREMA DE FEYNMAN-H-PAULI, PUES EL UNICO TERMINO QUE VARIA EN EL HAMILTONIANO TOTAL CON B ES EL DE LA ENERGIA MAGNETICA
Regarding energetics, the kinetic energy of the proton, $m_{p}\langle\mathbf{v}_{p}\rangle_{\mathbf{B}_{0}}^{2}/2=\langle\mathbf{P}_{N}\rangle_{\mathbf{B}_{0}}^{2}/2m_{p}$, is an $\mathcal{O}(e^{2}B^{2}_{0}/m^{3}_{p})$ correction to its EW self-energy. In Appendix \ref{AppE}, $\langle W(t_{0})\rangle$ is computed out of the integration of $\delta\langle W\rangle_{\mathbf{B}}$ from $\mathbf{B}=\mathbf{0}$ to $\mathbf{B}=\mathbf{B}_{0}$. There, it is shown that $\delta\langle W\rangle_{\mathbf{B}}$ contains a term 
\begin{equation}
-\langle\mathbf{P}_{N}\rangle_{\mathbf{B}}\cdot\delta\langle\mathbf{K}_{e^{+}}+\mathbf{Q}_{\nu}\rangle_{\mathbf{B}}/m_{p},\label{Doppler1}
\end{equation}
which is nothing but the energy associated to the differential Doppler-shift experienced  by the virtual leptons emitted and reabsorbed by the proton in motion along $\mathbf{B}$. The integration of Eq.(\ref{Doppler1}) leads straightaway to $\langle\mathbf{P}_{N}\rangle_{\mathbf{B}_{0}}^{2}/2m_{p}$.

Next, the question arises of what is the source of this energy.  It is immediate to prove that (see Appendix \ref{AppE}) 
\begin{equation}
\frac{\delta\langle W\rangle_{\mathbf{B}}}{\delta\mathbf{B}}\quad\textrm{is part of}\quad%\frac{\delta\langle W\rangle_{\mathbf{B}}}{\delta\mathbf{B}}\:\textrm{which is part of}
\left\langle\frac{\delta H_{mag}^{p}+\delta H_{mag}^{e}+\delta H_{mag}^{n}}{\delta\mathbf{B}}\right\rangle_{\mathbf{B}},\nonumber
\end{equation}
and hence, so is $\frac{\delta\langle\mathbf{P}_{N}\rangle_{\mathbf{B}}^{2}}{2m_{p}\delta\mathbf{B}}$. Therefore, we conclude that the kinetic energy of the proton is indeed a magnetic energy. This means that, while the kinetic momentum of the proton along $\mathbf{B}$ is provided by the virtual leptons, its kinetic energy is supplied by the source of the magnetic field. Thus, we interpret that the magnetic energy is spent in breaking the P and T symmetries which originates the asymmetry on the kinetic momentum of the virtual leptons along $\mathbf{B}.$

%Note also that the kinetic energy of the 0-Landau level of the proton is also provided by the magnetic source. However, the corresponding work on the $\hat{\mathbf{x}}\hat{\mathbf{y}}$-plane is done by the curly electric field instead, $\nabla\times\mathbf{E}=-\partial\mathbf{B}/\partial t$.

%That is, we can write $\langle\mathbf{P}_{N}\rangle_{\mathbf{B}_{0}}^{2}/2m_{p}=\beta_{EW}\mathbf{B}_{0}^{2}/2$, with $\beta_{EW}=|\delta\langle\mathbf{K}_{e}+\mathbf{Q}_{\nu}\rangle/\delta\mathbf{B}|^{2}/m_{p}$ being the EW-quantum correction to the magnetic polarisability of the proton.  

\section{Spin-relaxation-induced proton acceleration at constant magnetic field}\label{spinrelax}

%Let us consider next that the magnetic field remains constant in time at $\mathbf{B}_{0}$ for $t>t_{0}$. 
In the previous Section it was assumed implicitly that the state of the system $|\Phi\rangle_{\mathbf{B}}$, made of the superposition of the two lowest Landau levels of the proton with opposite spins, was a quasi-stationary state. This is so as long as the interaction of the proton with the electromagnetic vacuum field can be neglected. The latter is indeed a good approximation if the time of observation is much less than the inverse of the radiative emission rate, $\Gamma_{p}$, which makes the proton decay into its lowest Landau level in the presence of the magnetic field. That explains the condition $t\gtrsim t_{0}$, $t_{0}\ll\Gamma_{p}^{-1}$, in all the equations of the previous Section. 

In this Section we begin with the unpolarized proton state $|\Phi\rangle_{\mathbf{B}_{0}}$ at time $t_{0}$, and consider that the magnetic field remains constant in time at $\mathbf{B}_{0}$ for $t\geq t_{0}$. For simplicity, we neglect also the velocity of the proton at time $t_{0}$ given in Eq.(\ref{f1b}). In contrast to the previous Section, we incorporate the coupling between the EM vacuum and the proton through the Hamiltonian $H_{mag}^{p}$ by adding up the quantum field operator $\mathbf{A}$ to the uniform external field $\tilde{\mathbf{A}}$ considered already in Eq.(\ref{Hmag}). The Hamiltonian of the free EM field has the usual form,
\begin{equation}
H_{EM}=\frac{1}{2}\int\textrm{d}^{3}R[(\partial_{t}\mathbf{A})^{2}+(\nabla\times\mathbf{A})^{2}].
\end{equation}
The coupling of $\mathbf{A}$ to the proton makes the excited Landau levels decay through EM radiation. In particular, the coupling of the  quantum magnetic field $\nabla\times\mathbf{A}$ in $H_{mag}^{p}$ to the magnetic moment operator associated with the proton spin, causes the M1 transition from the state $|0,+1;\mathbf{0}\rangle^{p}_{\mathbf{B}_{0}}$  to $|0,-1;\mathbf{0}\rangle^{p}_{\mathbf{B}_{0}}$ in a finite time $\sim\Gamma_{p}^{-1}$. 
 In short, we write the effective Hamiltonian of the EM interaction as 
\begin{equation}
W_{mag}^{p}=-(2+\kappa_{p})\mu_{N}\mathbf{S}_{p}\cdot\nabla\times\mathbf{A},
\end{equation}
with $\mathbf{S}_{p}$ being the proton spin operator. In the following, we show that an additional Casimir momentum arises through the spin-relaxation of the proton.

\begin{figure}[h]
\includegraphics[height=7.0cm,width=8.3cm,clip]{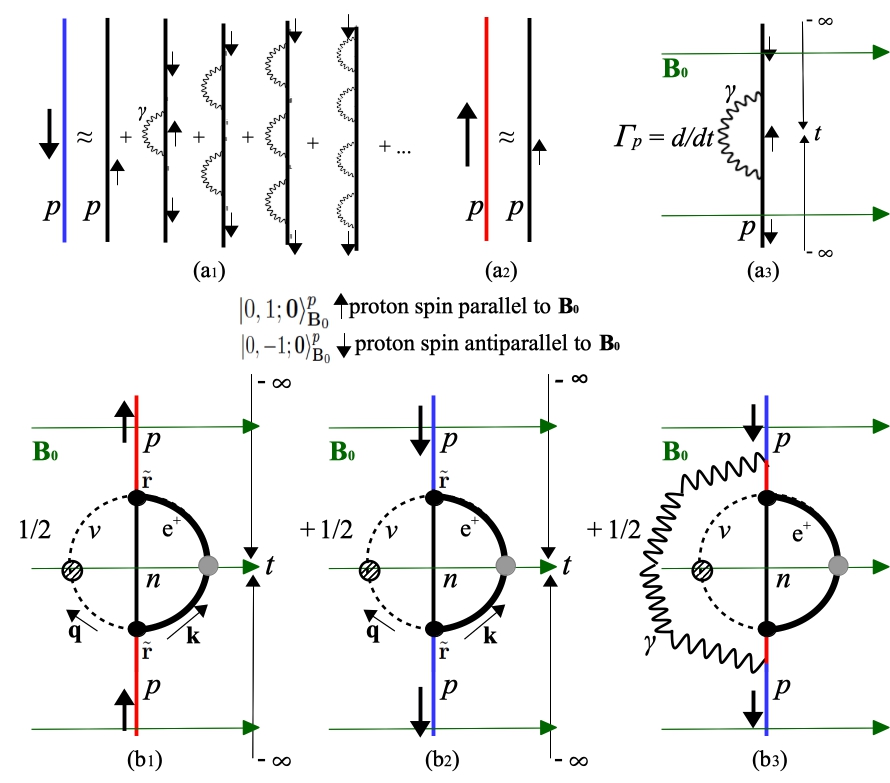}
\caption{(a) Diagrammatic representation of the spin-down proton state dressed by one-photon emission processes (a$_{1}$); the spin-up proton state in which EM vacuum effects are discarded (a$_{2}$); the one-photon emission rate $\Gamma_{p}$  according to Eq.(\ref{22}), where the proton is assumed adiabatically excited (a$_{3}$).  (b) Diagrammatic representation of $\langle(\mathbf{K}_{e^{+}}+\mathbf{Q}_{\nu})(t)\rangle$  according to Eqs.(\ref{equationmother2}) assuming the adiabatic excitation of the proton. Each diagram corresponds to the contribution of the first term of $|\Phi(t)\rangle$ on the \emph{right hand side (rhs)} of Eq.(\ref{23}) (b$_{1}$);  the third term of $|\Phi(t)\rangle$ on the \emph{rhs} of Eq.(\ref{23}) (b$_{2}$); and the second term of $|\Phi(t)\rangle$ on the \emph{rhs} of Eq.(\ref{23}) (b$_{3}$), respectively. The magnetic field remains being $\mathbf{B}_{0}$ at all time.}\label{fig2} %(b)\&(c) Schematic representation of the P and T symmetry-breaking processes in the transfer of momentum between the virtual leptons and the nucleon according to Eqs.(\ref{f1a})\&(\ref{f2a}), respectively. (d) Schematic representation of the nucleon-leptonic vacuum system considered as a whole, with null total momentum along $\mathbf{B}_{0}$.
\end{figure}

Let us start by defining time-propagators analogous to those of Sec.\ref{Dyna} which incorporate the Hamiltonians $H_{EM}$ and $W_{mag}^{p}$ in the adiabatic approximation. In the first place, we define $\mathbb{U}_{0}(\tau)=e^{-i[H_{0}(\mathbf{B}_{0})+H_{EM}]\tau}$,  and consider in $\tilde{\mathbb{U}}_{0}(t)$ the interaction of the proton with the EM vacuum field in the adiabatic approximation,
\begin{align}
\tilde{\mathbb{U}}_{0}(t)&=\mathbb{U}_{0}(t)\:T\exp{\left[-i\int_{-\infty}^{t}\textrm{d}\tau\:\mathbb{U}_{0}^{\dagger}(\tau)W_{mag}^{p}\mathbb{U}_{0}(\tau)e^{\eta\tau}\right]},\nonumber\\&\eta\rightarrow0^{+}.\label{Uad}
\end{align}
The adiabatic approximation in this equation is a consequence of the adiabatic evolution of the system along the adiabatic variation of the magnetic field up to its final value $\mathbf{B}_{0}$. Under the action of the EM quantum field, the state of the system evolves as $|\Phi(t)\rangle\simeq\tilde{\mathbb{U}}_{0}(t)[|\Phi\rangle_{\mathbf{B}_{0}}\otimes|\Omega_{\gamma}\rangle]$ for $t\gg t_{0}\gg\Gamma_{p}^{-1}$, with $|\Omega_{\gamma}\rangle$ being the EM vacuum state \footnote{It must be noted that the time interval of integration in  Eq.(\ref{Uad}) for $\tilde{\mathbb{U}}_{0}(t)$ is ill-defined in the adiabatic approximation. Therefore, it must be intended that only the finite leading order terms of its expansion in powers of $W_{mag}^{p}$ must be kept.}.  It is of note that the dominant contribution of $W_{mag}^{p}$ to  $|\Phi(t)\rangle$ is given by those processes which involve one-photon emission only from the proton state $|0,-1;\mathbf{0}\rangle^{p}_{\mathbf{B}_{0}}$. They are depicted in Fig.\ref{fig2}(a$_{1})$. They lead to an exponential decay in time whose rate is given by \cite{Buhmann,MelroseAstronAstroph1981,PothekinMonRoySoc2007}
\begin{align}
\Gamma_{p}&=2\textrm{Re}\:_{\mathbf{B}_{0}}^{p}\langle0,-1;\mathbf{0}|\otimes\langle\Omega_{\gamma}|\mathbb{U}_{0}^{\dagger}(t)\int_{-\infty}^{t}\textrm{d}\tau\:e^{\eta\tau}\mathbb{U}_{0}(t-\tau)\nonumber\\&\times W_{mag}^{p}\mathbb{U}_{0}(\tau)|\Omega_{\gamma}\rangle\otimes|0,-1;\mathbf{0}\rangle^{p}_{\mathbf{B}_{0}}
\simeq\mu_{N}^{5}\frac{(2+\kappa_{p})^{5}}{6\pi}B_{0}^{3},\nonumber\\&\quad\eta\rightarrow0^{+},\label{22}
\end{align}
whose diagram is depicted in Fig.\ref{fig2}(a$_{3})$. Finally, the effect of $W_{mag}^{p}$ on the ground state $|0,+1;\mathbf{0}\rangle^{p}_{\mathbf{B}_{0}}$ is a small shift in its magnetic energy that can be discarded --Fig.\ref{fig2}(a$_{2})$. Therefore, the net effect of the EM quantum field on the system is the spin-relaxation of the proton to its ground state. That is, in good approximation we can write 
\begin{align}
|\Phi(t)\rangle&=|\Omega_{l}\rangle_{\mathbf{B}_{0}}\otimes|\phi_{int}^{0}\rangle\otimes\Bigl[\frac{1}{\sqrt{E^{p}_{+}}}|0,1;\mathbf{0}\rangle^{p}_{\mathbf{B}_{0}}\otimes\Bigl(|\Omega_{\gamma}\rangle\nonumber\\
&+\frac{(1-e^{-\Gamma_{p}t})^{1/2}}{\sqrt{2k_{0}}L^{3/2}}\sum_{\hat{\mathbf{k}}_{0},s}|\mathbf{k}_{0};s\rangle{\gamma}\Bigr)\nonumber\\
&+\frac{e^{-\Gamma_{p}t/2}}{\sqrt{E^{p}_{-}}}|0,-1;\mathbf{0}\rangle^{p}_{\mathbf{B}_{0}}\otimes|\Omega_{\gamma}\rangle\Bigr]/2L,\label{23}
\end{align}
where $|\mathbf{k}_{0};s\rangle{\gamma}/\sqrt{2k_{0}L^{3}}$ is the normalized state of a photon of momentum $\mathbf{k}_{0}$ and polarisation $s$ (see Appendix \ref{AppA4}), with $k_{0}=\mu_{N}B_{0}(2+\kappa_{p})$. 

Lastly, the EW interaction remains perturbative. At lowest order in $W$, the total time-propagator is now
\begin{equation}
\tilde{\mathbb{U}}(t)\equiv\tilde{\mathbb{U}}_{0}(t)-i\int_{-\infty}^{t}\textrm{d}\tau\:\tilde{\mathbb{U}}_{0}(t-\tau)W\tilde{\mathbb{U}}_{0}(\tau)e^{\eta\tau},\:\eta\rightarrow0^{+}.\label{eta2}
\end{equation}

As spin-relaxation takes place,  it is symmetry allowed for virtual positrons and neutrinos to acquire a nonvanishing momentum along the proton spin.  Using Eq.(\ref{eta2}) we write this momentum as
\begin{align}
\langle(\mathbf{K}_{e^{+}}&+\mathbf{Q}_{\nu})(t)\rangle=\:_{\mathbf{B}_{0}}\langle\Phi|\:\tilde{\mathbb{U}}^{\dagger}(t)(\mathbf{K}_{e^{+}}+\mathbf{Q}_{\nu})\tilde{\mathbb{U}}(t)\:|\Phi\rangle_{\mathbf{B}_{0}}\nonumber\\
&\simeq\frac{G_{F}^{2}}{2}\hat{\mathbf{z}}\int\textrm{d}^{3}R\textrm{d}^{3}R'\:\langle\Phi(t)|J_{h}^{\mu}(\tilde{\mathbf{r}})J^{l}_{\mu}(\tilde{\mathbf{r}})\label{equationmother2}\\&\times\frac{(\mathbf{K}_{e^{+}}+\mathbf{Q}_{\nu})\cdot\hat{\mathbf{z}}}{[H_{0}(\mathbf{B}_{0})-E_{0}^{p}]^{2}}J^{\rho\dagger}_{\mu}(\tilde{\mathbf{r}}')J_{h}^{\rho\dagger}(\tilde{\mathbf{r}}')|\Phi(t)\rangle,\quad t\gg t_{0}.\nonumber
\end{align}
This equation is identical to Eq.(\ref{equationmother}) but for the replacement of $|\Phi\rangle_{\mathbf{B}_{0}}$ with $|\Phi(t)\rangle$. Its diagrammatic representation is that of Figs.\ref{fig2}(b$_{1}$-b$_{3})$ %The energy of photon radiated during the spin-flip process, $k_{0}$, has been discarded in the denominator of the integrand.  
 It is of note that the expectation value of the momentum of the emitted photon is identically zero at leading order. It is only when considering the proton velocity that the differential Doppler-shift experienced by the photon along $\langle\mathbf{P}_{N}\rangle$ results in directional emission. Therefore, at our order of approximation it holds that $\langle\mathbf{P}_{N}(t)\rangle=-\langle(\mathbf{K}_{e^{+}}+\mathbf{Q}_{\nu})(t)\rangle$ and we can ignore the photon-emission-induced recoil. Taking into account these considerations we arrive at 
\begin{widetext}
\begin{align}
&\langle(\mathbf{K}_{e^{+}}+\mathbf{Q}_{\nu})(t)\rangle=\simeq\textrm{Re}\int\textrm{d}y\:\textrm{d}y'\int\frac{\textrm{d}p_{y}}{2\pi}\int\frac{\textrm{d}^{3}q}{(2\pi)^{3}}\int\frac{\textrm{d}k_{x}\textrm{d}k_{z}}{(2\pi)^{2}}e^{i(p_{y}+q_{y})(y-y')}(k_{z}+q_{z})\hat{\mathbf{z}}\sum_{\mathcal{N},m=0}^{\infty}G_{F}^{2}\Bigg\{\frac{4I_{0}(\xi_{p})I_{\mathcal{N}}(\xi_{e^{+}})}{q\mathcal{E}_{\mathcal{N},m}^{2}E^{e}_{\mathcal{N}}}\nonumber\\
&\times I_{0}(\xi^{'}_{p})I_{\mathcal{N}}(\xi^{'}_{e^{+}})\Bigg( \mathbf{k}\cdot\langle\mathbf{S}_{p}(t)\rangle\Big[q(g^{2}_{A}-g_{A})-\frac{\mathbf{q}\cdot\mathbf{p}}{2m_{n}}(g_{A}+1)^{2}\Big]
+\mathbf{q}\cdot\langle\mathbf{S}_{p}(t')\rangle E^{e}_{\mathcal{N}}(g_{A}^{2}+g_{A})+\mathbf{p}\cdot\langle\mathbf{S}_{p}(t)\rangle\frac{\mathbf{q}\cdot\mathbf{k}}{2m_{n}}(g_{A}-1)^{2}\Bigg)\nonumber\\
&+\frac{2\sqrt{2NeB_{0}}I_{\mathcal{N}}(\xi_{e^{+}})I_{0}(\xi_{p})I_{N-1}(\xi^{'}_{e^{+}})I_{0}(\xi^{'}_{p})}{m_{n}q\mathcal{E}_{\mathcal{N},m}^{2}E^{e}_{\mathcal{N}}}\mathbf{q}\cdot\langle\mathbf{S}_{p}(t)\rangle\frac{\mathbf{p}\cdot\mathbf{A}}{A}(g_{A}-1)^{2}\Bigg\}\Bigl|\langle\tilde{\phi}_{int}^{0}|\tilde{\phi}_{int}^{m}\rangle\Bigr|^{2},\label{f2a}
\end{align}
\end{widetext}
where $\langle\mathbf{S}_{p}(t)\rangle=(1-e^{-\Gamma_{p}t})\hat{\mathbf{z}}/2$ is the expectation value of the proton spin. Truncating the sum over internal nucleon states at $m=0$ as for Eq.(\ref{f1b}), %Refs.\cite{Coince,IsgurKarl1978,IsgurKarl1979}, 
the integration of Eq.(\ref{f2a}) yields
%\begin{align} 
%&\langle\mathbf{K}_{e}(t)+\mathbf{Q}_{\nu}(t)\rangle\approx\frac{G_{F}^{2}m_{d}^{5}}{8\pi^{4}}\langle\mathbf{S}_{p}(t)\rangle\Big[1.7g_{A}^{2}+0.5g_{A}\nonumber\\
%&\qquad\qquad\qquad\:\:+\frac{m_{d}}{m_{n}}(6g_{A}-g_{A}^{2}-1)\Big],\label{f2b}\\
%&\langle\mathbf{v}_{p}(t)\rangle\approx-\mu\textrm{m/s}\:(1-e^{-\Gamma_{p}t})\:\hat{\mathbf{z}}.\label{f2bp}
%\end{align} 
\begin{align} 
&\langle\mathbf{K}_{e^{+}}(t)+\mathbf{Q}_{\nu}(t)\rangle\approx\frac{G_{F}^{2}m_{d}^{5}}{8\pi^{4}}\langle\mathbf{S}_{p}(t)\rangle\Big[1.7g_{A}^{2}+0.5g_{A}+\frac{m_{d}}{m_{n}}\nonumber\\
&\times(6g_{A}-g_{A}^{2}-1)\Big],\:\:\langle\mathbf{v}_{p}(t)\rangle\approx-\frac{\mu\textrm{m}}{\textrm{s}}(1-e^{-\Gamma_{p}t})\:\hat{\mathbf{z}}.\label{f2b}
\end{align} 
Let us emphasize that the result of Eq.(\ref{f2b}) is independent of the process by means of which the proton was polarized, but for the specific value of $\Gamma_{p}$. At he end of the polarization process, the proton velocity is of the order of $\mu$m/s, which may lie within the scope of current experimental observations using proton traps \cite{Penning}.
  
%ONE MAY WONDER WHETHER THIS MOMENTUM COULD COME FROM THE RECOIL IN THE PHOTON EMISSION, WHICH CANNOT BE THE CASE BECAUSE PHOTON EMISSION IS ISOTROPIC AND SECOND, THE RECOIL WOULD DEPEND ON THE PARTICULAR VALUE OF Bo 

\section{Conclusions}\label{Conclusions}

We have shown that virtual leptons acquire a Casimir momentum when EW-coupled to a proton in the presence of a spatially uniform magnetic field. Reciprocally, an equivalent kinetic momentum is transferred to the proton in the opposite direction. In the first place, under the action of an adiabatically increasing magnetic field, $\mathbf{B}$, the resultant momentum is proportional  to the magnetic field at any time [Eq.(\ref{f1b})]. In addition, at constant magnetic field, as the spin-relaxation of the proton proceeds, the momentum is proportional to the expectation value of the proton spin, $\langle\mathbf{S}_{p}\rangle$ [Eq.(\ref{f2b})]. At the end of the spin-relaxation process, the proton velocity, antiparallel to $\langle\mathbf{S}_{p}\rangle$, is of the order of $\mu/$s, which may be experimentally accessible  and constitutes the most important result of this article.  

Therefore, contrary to the result in classical electrodynamics  \cite{Griffiths}, our findings in EW and those of Ref.\cite{PRLDonaire} in EM imply that, generically, a magnetic particle does accelerate linearly along a spatially uniform magnetic field if its interaction with the quantum vacuum is chiral. Further, in the EW case, there is no need for the particle to be magnetic (eg., an unstable neutral pion), since virtual leptons and hadrons can be spin-polarized too. Hence, since the EW interaction of any material particle with the leptonic and hadronic vacua is chiral, it follows that, in the presence of a magnetic field, the Casimir momentum is a generic phenomenon in nature.%That means
%or instance that a neutral pion, although unstable, could
%be accelerated slightly by a spatially uniform time-varying magnetic field.  %That means for instance  that a neutral pion, although unstable, could be accelerated by a uniform magnetic field.

Lastly, for the case that the proton accelerates under the action of an adiabatically time-varying magnetic field, it has been demonstrated that its kinetic energy is a magnetic energy. This means that, while the kinetic momentum of the proton along $-\mathbf{B}$ is provided by the virtual leptons, its kinetic energy is supplied by the source of the magnetic field. Also, since leptons remain virtual throughout the process, this implies that the proton acceleration is not caused by the recoil associated to the emission of any actual particle.

\section{Discussion}\label{Discussion}

Regarding the approach, some comments are in order. In the first place, the most promising result to be verified experimentally is that of the velocity of a polarized proton in Eq.(\ref{f2b}). However, spin-relaxation via EM emission, as presented in Sec.\ref{spinrelax}, is a highly inefficient method. Nonetheless, as long as adiabaticity is preserved in the polarization process, any other more efficient technique should yield an equivalent result \cite{PRL106_253001(2011),Penning}. Second, the results of Eqs.(\ref{f1b}) and (\ref{f2b}) depend on the hadron model which determines the nucleon wavefunction in Eqs.(\ref{f1a}) and (\ref{f2a}) \cite{Coince,IsgurKarl1978,IsgurKarl1979,deRujula,Isgur}. %and on the neglect of the quark magnetic excitations \cite{remark,Karliner}. 
%our results rest on the assumptions that quarks are nonrelativistic within the nucleon, and that the nucleon interacts with the magnetic field as a whole \cite{remark,Karliner}. The former  implies a cutoff of order $m_{d}$ in the momentum integrals, which we have accounted for using a Gaussian wavefunction for the active quark \cite{SM,Coince,IsgurKarl1978,IsgurKarl1979,QuarkModels}. The latter involves the neglect of magnetic quark excitations. 
Consequently, discrepancies of order unity are expected between our estimates and the experimental measurements. Nonetheless, this makes the leptonic Casimir momentum a suitable phenomenon to test hadron models at low energies \cite{cavitychromo,QuarkModels}. Improvements on the accuracy of our results involve the implementation of relativistic hadron models in the internal nucleon Hamiltonian \cite{Shulga}, as well as including relativistic corrections to the external dynamics of the intermediate neutron state. In any case, it is noteworthy that, as for the calculation of the Lamb-shift, the convergence of the resultant momentum integrals is guaranteed in our approach by the inclusion of retardation factors in the lepton fields, together with the internal wavefunction of the nucleon. 

\begin{figure}[h]
\includegraphics[height=4.1cm,width=4.cm,clip]{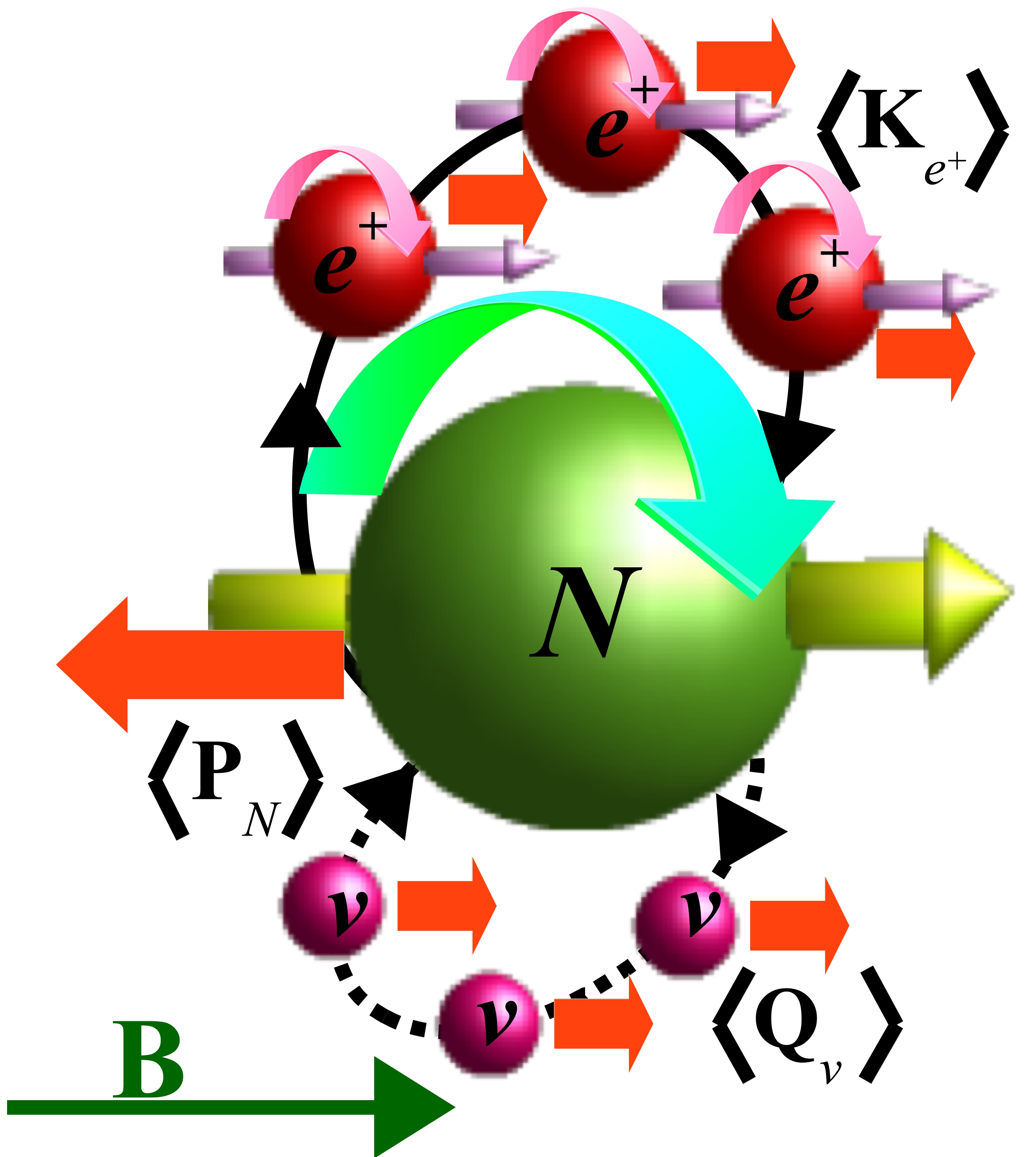}
\caption{Pictorial representation of total momentum conservation on the nucleon-leptonic-vacuum state considered as a whole, $\langle\mathbf{P}_{N}+\mathbf{K}_{e^{+}}+\mathbf{Q}_{\nu}\rangle=\mathbf{0}$.}\label{fig4}
\end{figure}

Concerning the physical interpretation of our findings, we first comment on the neglect of the photon momentum in Eq.(\ref{f2a}). It was argued in Sec.\ref{spinrelax} that the expectation value of the photon emitted in the spin-relaxation process is null. However, when considering a single emission realization --see Fig.\ref{fig2}($b_{3})$, the emitted photon carries a net momentum of modulus $k_{0}\sim B_{0}^{3}/m_{p}^{5}$ which causes the recoil of the proton. For laboratory magnetic fields, $k_{0}$ is several orders of magnitude smaller than the lepton Casimir momentum of Eq.(\ref{f2b}), and thus negligible in any single emission realization. Second, at a fundamental level, it is worth emphasizing that the Casimir momentum, as any other observable quantum phenomenon, is to be attributed to virtual particles or quantum field fluctuations coupled to actual particles. Its adscription to the vacuum, as appears  in Ref.\cite{JPCMDonaire} in an EM scenario, rests on the interpretation that the vacuum state and the state of the electric charges are disentangled, which is obviously not the case. As a matter of fact, in our EW scenario it is the consideration of the nucleon-leptonic-vacuum state as a whole which allows us to unravel the apparent violation of the action-reaction principle. That is, since no transfer of momentum with the source of magnetic field takes place at any time, it is the total momentum of the nucleon-leptonic-vacuum state which remains null, in accordance with momentum conservation and global translation invariance along $\mathbf{B}$. This is represented pictorially in Fig.\ref{fig4}.

Finally, let us discuss some aspects to be addressed in future work. In the first place, it is left to investigate the nature of the kinetic energy of the proton at the end of the spin-relaxation process. Interestingly, it seems not to depend on the strength of the magnetic field. Second, we plan to apply the formalism to the calculation of the acceleration of a neutron  as it gets spin-polarized. This calculation is relevant for two reasons. First, from an experimental perspective, in contrast to protons, it might be easier to test the acceleration of an ensemble of polarized neutrons which are not affected by the Coulomb interaction. Second, due to the neutron instability  against beta decay, it is expected that the momentum integrals analogous to those in Eq.(\ref{f2a}) possess a resonant component \cite{myPRA} in addition to the off-resonant one found for the proton. The question arises as to whether the well-known directionality in the emission of beta particles would be the result of a \emph{resonant  Casimir momentum} only or if, on the contrary, they would carry with them the \emph{total Casimir momentum} of virtual electrons.
 
%Let us finalize with a comment on the nature of the force $\langle\mathbf{F}\rangle_{\mathbf{B}}$ in Eq.(SM25). In the first place, it is obvious that this force is not a Lorentz force. Indeed, this force on the proton is the result of the unbalanced momentum induced on the virtual leptons of the vacuum by the joint action of the magnetic field in $H_{mag}^{\alpha}$, $\alpha=p,n,e$, and the chiral EW potential $W$. Therefore, that force has both a magnetic and an electroweak origin. Nevertheless, as has just been proved, the kinetic energy and hence the work performed by this force is provided only by the source of magnetic field.

\acknowledgments
We thank A. Cano, L. Froufe, M. Chernodub and B. van Tiggelen for fruitful discussions on the subject.
Financial support from grants MTM2014-57129-C2-1-P (MINECO), BU229P18 and VA137G18 (JCyL) is acknowledged.

\appendix
\begin{widetext}
\section{Spinor eigenstates of magnetic leptons and nucleons in a uniform and stationary magnetic field}\label{AppA}

In this Appendix we compile the equations for the spinor eigenstates of the Hamiltonian $H_{D}^{\alpha}+H_{mag}^{\alpha}$ for electrons ($\alpha=e$), protons ($\alpha=p$) and neutrons ($\alpha=n$) in the presence of an external and uniform magnetic field, $\mathbf{B}=B\hat{\mathbf{z}}$, in the gauge $\tilde{\mathbf{A}}=-Br_{y}\hat{\mathbf{x}}$. We make use of the results within Refs.\cite{Mao,Bhattacharya,Bander,Broderick}, as well as the textbooks of Peskin \& Schroeder and Itzykson \& Zube \cite{Peskin}. We employ the chiral representation, with the Dirac matrices given by
\[
  \gamma_{0}=
  \left[ {\begin{array}{cc}
  0 & - \mathbb{I}\\       -\mathbb{I}&0\      \end{array} } \right],\quad \gamma_{5}=
  \left[ {\begin{array}{cc}
    \mathbb{I} & 0\\       0 & -\mathbb{I}\      \end{array} } \right],\quad \gamma_{i}=
  \left[ {\begin{array}{cc}
    0 & \sigma_{i}\\       -\sigma_{i} &0\      \end{array} } \right],
\]
where $\mathbb{I}$  is the two-dimensional identity matrix and $\sigma_{i}$ is the $i^{th}$ Pauli matrix. In this representation, $\gamma_{5}$ writes $\gamma_{5}=+i\gamma_{0}\gamma_{1}\gamma_{2}\gamma_{3}$.

\subsection{Electron and positron Landau levels}\label{posit}

For electrons, the eigenstates of the Hamiltonian density operator, $\hat{h}_{D}^{e}+\hat{h}_{mag}^{e}=\gamma_{0}[-i\gamma^{j}\partial_{j}-eB\:r_{y}\gamma_{1}+m_{e}\mathbb{I}]$, $j=1,2,3$, correspond to the spinors of Landau levels with positive and negative energies, $E^{e}_{\mathcal{N}}=\sqrt{m_{e}^{2}+k_{z}^{2}+2\mathcal{N}eB}$, where $\mathcal{N}$ is the principal quantum number, $\mathcal{N}=0,1,2,3,...$, and $k_{z}$ is the momentum component along $\mathbf{B}$. Both electron and positron eigenstates are degenerate with respect to the spin number, $\mathcal{S}=\pm1$, except for the ground state $\mathcal{N}=0$ for which a unique solution exists with $\mathcal{S}=-1$.

The spinor eigenstates  for electrons read
\[
U^{e}_{\mathcal{N},+}=\left[ \begin{array}{c} \frac{E^{e}_{\mathcal{N}}+m_{e}+k_{z}}{\sqrt{2(E^{e}_{\mathcal{N}}+m_{e})}}I_{\mathcal{N}-1}(\xi_{e})\\\\-\sqrt{\frac{\mathcal{N}eB}{E^{e}_{\mathcal{N}}+m_{e}}}I_{\mathcal{N}}(\xi_{e})\\\\-\frac{E^{e}_{\mathcal{N}}+m_{e}-k_{z}}{\sqrt{2(E^{e}_{\mathcal{N}}+m_{e})}}I_{\mathcal{N}-1}(\xi_{e})\\\\-\sqrt{\frac{\mathcal{N}eB}{E^{e}_{\mathcal{N}}+m_{e}}}I_{\mathcal{N}}(\xi_{e})\end{array} \right],\quad
  U^{e}_{\mathcal{N},-}=\left[ \begin{array}{c}-\sqrt{\frac{\mathcal{N}eB}{E^{e}_{\mathcal{N}}+m_{e}}}I_{\mathcal{N}-1}(\xi_{e})\\\\\frac{E^{e}_{\mathcal{N}}+m_{e}-k_{z}}{\sqrt{2(E^{e}_{\mathcal{N}}+m_{e})}}I_{\mathcal{N}}(\xi_{e})\\\\-\sqrt{\frac{\mathcal{N}eB}{E^{e}_{\mathcal{N}}+m_{e}}}I_{\mathcal{N}-1}(\xi_{e})\\\\-\frac{E^{e}_{\mathcal{N}}+m_{e}+k_{z}}{\sqrt{2(E^{e}_{\mathcal{N}}+m_{e})}}I_{\mathcal{N}}(\xi_{e})\end{array} \right], \textrm{ for }\mathcal{S}=\pm1\textrm{ respectively, } \xi_{e}=\sqrt{eB}(r_{y}-\frac{k_{x}}{eB});
\]
 while those for positrons are   
\[
V^{e}_{\mathcal{N},+}=\left[ \begin{array}{c}\frac{E^{e}_{\mathcal{N}}+m_{e}+k_{z}}{\sqrt{2(E^{e}_{\mathcal{N}}+m_{e})}}I_{\mathcal{N}-1}(\xi_{e^{+}})\\\\\sqrt{\frac{\mathcal{N}eB}{E^{e}_{\mathcal{N}}+m_{e}}}I_{\mathcal{N}}(\xi_{e^{+}})\\\\\frac{E^{e}_{\mathcal{N}}+m_{e}-k_{z}}{\sqrt{2(E^{e}_{\mathcal{N}}+m_{e})}}I_{\mathcal{N}-1}(\xi_{e^{+}})\\\\-\sqrt{\frac{\mathcal{N}eB}{E^{e}_{\mathcal{N}}+m_{e}}}I_{\mathcal{N}}(\xi_{e^{+}})\end{array} \right],\quad
  V^{e}_{\mathcal{N},-}=\left[ \begin{array}{c} \sqrt{\frac{\mathcal{N}eB}{E^{e}_{\mathcal{N}}+m_{e}}}I_{\mathcal{N}-1}(\xi_{e^{+}})\\\\\frac{E^{e}_{\mathcal{N}}+m_{e}-k_{z}}{\sqrt{2(E^{e}_{\mathcal{N}}+m_{e})}}I_{\mathcal{N}}(\xi_{e^{+}})\\\\-\sqrt{\frac{\mathcal{N}eB}{E^{e}_{\mathcal{N}}+m_{e}}}I_{\mathcal{N}-1}(\xi_{e^{+}})\\\\\frac{E^{e}_{\mathcal{N}}+m_{e}+k_{z}}{\sqrt{2(E^{e}_{\mathcal{N}}+m_{e})}}I_{\mathcal{N}}(\xi_{e^{+}})\end{array} \right], \textrm{ for }\mathcal{S}=\pm1\textrm{ respectively, } \xi_{e^{+}}=\sqrt{eB}(r_{y}+\frac{k_{x}}{eB}),
\]
such that $\hat{h}_{D}^{e}e^{ik_{x}r_{x}+ik_{z}r_{z}}U^{e}_{\mathcal{N},\pm}=E^{e}_{\mathcal{N}}e^{ik_{x}r_{x}+ik_{z}r_{z}}U^{e}_{\mathcal{N},\pm}$, $\hat{h}_{D}^{e}e^{-ik_{x}r_{x}-ik_{z}r_{z}}V^{e}_{\mathcal{N},\pm}=-E^{e}_{\mathcal{N}}e^{-ik_{x}r_{x}-ik_{z}r_{z}}V^{e}_{\mathcal{N},\pm}$.    The functions $I_{\mathcal{N}}(X)$ were defined in Sec.\ref{Adiamomentum} as $I_{\mathcal{N}}(X)=(eB)^{1/4}H_{\mathcal{N}}(X)e^{-X^{2}/2}/\pi^{1/4}\sqrt{2^{\mathcal{N}}\mathcal{N}!}$, where $H_{\mathcal{N}}$ is the Hermite polynomial of order $\mathcal{N}$. In terms of the above eigenstates, the Dirac field of electrons reads, in Schr\"odinger's representation,
\begin{equation}
\Psi_{e}(\mathbf{r})=\sum_{\mathcal{S}=\pm1}\sum_{\mathcal{N}=0}^{\infty}\int\frac{\textrm{d}k_{x}\textrm{d}k_{z}}{(2\pi)^{2}\sqrt{2E^{e}_{\mathcal{N}}}}\bigl[a^{e}_{\mathcal{N},\mathcal{S}}(k_{x},k_{z})\:U^{e}_{\mathcal{N},\mathcal{S}}\:e^{i(k_{x}r_{x}+k_{z}r_{z})}+b^{e\dagger}_{\mathcal{N},\mathcal{S}}(k_{x},k_{z})\:V^{e}_{\mathcal{N},\mathcal{S}}\:e^{-i(k_{x}r_{x}+k_{z}r_{z})}\bigr],\label{SM1}
\end{equation}
where $a^{e(\dagger)}_{\mathcal{N},\mathcal{S}}(k_{x},k_{z})$ and $b^{e(\dagger)}_{\mathcal{N},\mathcal{S}}(k_{x},k_{z})$ are the annihilation and creation ($\dagger$) operators of electrons and positrons, respectively, which satisfy the usual anticommutation relations,
\begin{equation}
\{a^{e}_{\mathcal{N},\mathcal{S}}(k_{x},k_{z}),a^{e\dagger}_{\mathcal{N}',\mathcal{S}'}(k_{x}',k_{z}')\}=\{b^{e}_{\mathcal{N},\mathcal{S}}(k_{x},k_{z}),b^{e\dagger}_{\mathcal{N}',\mathcal{S}'}(k_{x}',k_{z}')\}=(2\pi)^{2}\delta(k_{x}-k_{x}')\delta(k_{z}-k_{z}')\delta_{\mathcal{N}\mathcal{N}'}\delta_{\mathcal{S}\mathcal{S}'}.\label{SM2}
\end{equation}
Positron states appear as vacuum fluctuations in the calculations of the Letter. In terms of positron creation operators acting on the leptonic vacuum, their expression is
\begin{equation}
|\mathcal{N},\mathcal{S};k_{x},k_{z}\rangle_{\mathbf{B}}^{e}=\sqrt{2E_{\mathcal{N}}^{e}}\:b_{\mathcal{N},\mathcal{S}}^{e\dagger}(k_{x},k_{z})|\Omega_{l}\rangle_{\mathbf{B}}.\nonumber
\end{equation}

\subsection{Low-lying proton Landau levels}

As for protons, we will restrict ourselves to the low-lying positive energy eigenstates of the Hamiltonian operator $\hat{h}_{D}^{p}+\hat{h}_{mag}^{p}=\gamma_{0}\bigl[-i\gamma^{j}\partial_{j}+eB\:r_{y}\gamma_{1}+m_{p}\mathbb{I}+i\kappa_{p}\mu_{N}B[\gamma_{2},\gamma_{1}]/4\bigr]$, $j=1,2$. These are, the positive energy spinors corresponding to the Landau levels with kinetic momentum  $p_{z}$ along $\mathbf{B}$, $|p_{z}|\ll\sqrt{eB}\ll m_{p}$, principal number $\mathcal{N}=0$ and spin numbers $S=+1$ and $S=-1$, with energies $E^{p}_{+}\simeq m_{p}+p_{z}^{2}/2m_{p}-\kappa_{p}eB/4m_{p}$ and $E^{p}_{-}\simeq m_{p}+p_{z}^{2}/2m_{p}+\kappa_{p}eB/4m_{p}+eB/m_{p}$, respectively,
\[
U^{p}_{0,+}\simeq i\sqrt{E^{p}_{+}}I_{0}(\xi_{p})\left[ \begin{array}{c}1\\\\0\\\\\ -1\\\\0\end{array} \right],\quad
  U^{p}_{0,-}\simeq (1-\frac{eB}{4m_{p}^{2}})\sqrt{E^{p}_{-}}\left[ \begin{array}{c}-i\frac{\sqrt{eB}}{\sqrt{2}m_{p}}I_{1}(\xi_{p})\\\\-I_{0}(\xi_{p})\\\\-i\frac{\sqrt{eB}}{\sqrt{2}m_{p}}I_{1}(\xi_{p})\\\\I_{0}(\xi_{p})\end{array} \right], \textrm{ for }\mathcal{S}=\pm1,\textrm{  respectively, } \xi_{p}=\sqrt{eB}(r_{y}+\frac{p_{x}}{eB}),
\]
such that $(\hat{h}_{D}^{p}+\hat{h}_{mag}^{p})\:e^{i(p_{x}r_{x}+p_{z}r_{z})}U^{p}_{0,\pm}=E^{p}_{\pm}e^{i(p_{x}r_{x}+p_{z}r_{z})}U^{p}_{0,\pm}$.  

Creation and annihilation proton (and antiproton) operators follow anticommutation relations analogous to those in Eq.(\ref{SM2}),
\begin{equation}
\{a^{p}_{\mathcal{N},\mathcal{S}}(p_{x},p_{z}),a^{p\dagger}_{\mathcal{N}',\mathcal{S}'}(p_{x}',p_{z}')\}=\{b^{p}_{\mathcal{N},\mathcal{S}}(p_{x},p_{z}),b^{p\dagger}_{\mathcal{N}',\mathcal{S}'}(p_{x}',p_{z}')\}=(2\pi)^{2}\delta(p_{x}-p_{x}')\delta(p_{z}-p_{z}')\delta_{\mathcal{N}\mathcal{N}'}\delta_{\mathcal{S}\mathcal{S}'},\nonumber
\end{equation}
and the proton field reads
\begin{equation}
\Psi_{p}(\mathbf{r})=\sum_{\mathcal{S}=\pm1}\sum_{\mathcal{N}=0}^{\infty}\int\frac{\textrm{d}p_{x}\textrm{d}p_{z}}{(2\pi)^{2}\sqrt{2E^{p}_{\mathcal{N},\mathcal{S}}}}\bigl[a^{p}_{\mathcal{N},\mathcal{S}}(p_{x},p_{z})\:U^{p}_{\mathcal{N},\mathcal{S}}\:e^{i(p_{x}r_{x}+p_{z}r_{z})}+b^{p\dagger}_{\mathcal{N},\mathcal{S}}(p_{x},p_{z})\:V^{p}_{\mathcal{N},\mathcal{S}}\:e^{-i(p_{x}r_{x}+p_{z}r_{z})}\bigr],\nonumber
\end{equation}
where $V^{p}_{\mathcal{N},\mathcal{S}}$ is the antiproton eigenstate and $E^{p}_{\mathcal{N},\mathcal{S}}$ the proton energy of the Landau level $\mathcal{N}$ with spin number $\mathcal{S}$. Relevant to our calculations are only the expressions for $U^{p}_{0,+}$ and $U^{p}_{0,-}$ given above.  As for the minimum energy external state of an unpolarized proton in the presence of a magnetic field $\mathbf{B}$, it is given in Eq.(\ref{PB}) of Sec.\ref{Dyna} by $|\phi_{ext}^{p}\rangle_{\mathbf{B}}=\bigl[|0,1;\mathbf{p}\rangle+|0,-1;\mathbf{p}\rangle\bigr]^{p}_{\mathbf{B}}/2L\sqrt{E_{0}^{p}}$, with $E_{0}^{p}=(E_{+}^{p}+E_{-}^{p})/2$. In terms of proton creation operators acting on the hadronic vacuum, $|\Omega_{h}\rangle_{\mathbf{B}}$, the two Landau levels are 
\begin{equation}
|0,1;\mathbf{p}\rangle^{p}_{\mathbf{B}}=\sqrt{2E_{+}^{p}}a^{p\dagger}_{0,1}(p_{x},p_{z})|\Omega_{h}\rangle_{\mathbf{B}},\quad|0,-1;\mathbf{p}\rangle^{p}_{\mathbf{B}}=\sqrt{2E_{-}^{p}}a^{p\dagger}_{0,-1}(p_{x},p_{z})|\Omega_{h}\rangle_{\mathbf{B}}.\nonumber
\end{equation}

It is worth mentioning that the kinetic energy of the 0-Landau levels of the proton, related to its motion on the plane perpendicular to $\mathbf{B}$, is  provided by the work  done by the classical rotational electric field, $\nabla\times\mathbf{E}=-\partial\mathbf{B}/\partial\tau$. 

\subsection{Nonrelativistic neutron states and massless neutrino states}

Neutrons in the presence of a magnetic field $\mathbf{B}$ experience Zeeman splitting as a result of their anomalous magnetic moment. The eigenvalues of $\hat{h}_{D}^{n}+\hat{h}_{mag}^{n}=\gamma_{0}\bigl[-i\gamma^{j}\partial_{j}+m_{n}\mathbb{I}+i\kappa_{n}\mu_{N}B[\gamma_{2},\gamma_{1}]/4\bigr]$, $j=1,2,3$, for nonrelativistic neutrons of kinetic momentum $p\ll m_{n}$, and $eB\ll m_{n}^{2}$, depend on the quantum spin number, $\mathcal{S}=\pm1$, as $E^{n}_{\mathcal{S}}\simeq m_{n} + p^{2}/2m_{n}+\mathcal{S}|\kappa_{n}|\mu_{N}B/2$. In the  nonrelativistic limit, at leading order in $eB/m_{n}^{2}$, the neutron eigenstates are

\[
U^{n}_{\mathcal{S}}=\sqrt{E_{\mathcal{S}}^{n}}\frac{(1+p_{z}/2m_{n}-\mathcal{S}p_{z}|g_{n}|\mu_{N}B/4m^{2}_{n})(p_{x}+ip_{y})}{(\mathcal{S}+1)m_{n}+p_{x}+ip_{y}}\left[ \begin{array}{c}\frac{(\mathcal{S}+1)m_{n}+\mathcal{S}|p_{x}+ip_{y}|^2/2m_{n}}{p_{x}+ip_{y}}\\\\\mathcal{S}(1-p_{z}/m_{n})\\\\-\mathcal{S}(1-p_{z}/m_{n})\frac{(\mathcal{S}+1)m_{n}+\mathcal{S}|p_{x}+ip_{y}|^2/2m_{n}}{p_{x}+ip_{y}}\\\\1\end{array} \right]
\]
\[
+\sqrt{E_{\mathcal{S}}^{n}}\frac{(1+p_{z}/2m_{n})(p_{x}+ip_{y})|g_{n}|\mu_{N}B}{2(\mathcal{S}+1)m_{n}+p_{x}+ip_{y}}\left[ \begin{array}{c}0\\\\p_{z}/m_{n}^{2}\\\\-(\mathcal{S}+1)\frac{1+p_{z}/m_{n}}{p_{x}+ip_{y}}\\\\0\end{array} \right], \textrm{ for }\mathcal{S}=\pm1,
\] 
such that $(\hat{h}_{D}^{n}+\hat{h}_{mag}^{n})e^{i\mathbf{p}\cdot\mathbf{r}}U^{n}_{\mathcal{S}}=E_{\mathcal{S}}^{n}e^{i\mathbf{p}\cdot\mathbf{r}}U^{n}_{\mathcal{S}}$. The neutron field $\Psi_{n}(\mathbf{r})$ reads 
\begin{equation}
\Psi_{n}(\mathbf{r})=\sum_{\mathcal{S}=\pm1}\int\frac{\textrm{d}^{3}p}{(2\pi)^{3}}\frac{1}{\sqrt{2E^{n}_{\mathcal{S}}}}\bigl[a^{n}_{\mathcal{S}}(\mathbf{p})\:U^{n}_{\mathbf{p},\mathcal{S}}\:e^{i\mathbf{p}\cdot\mathbf{r}}+b^{n\dagger}_{\mathcal{S}}(\mathbf{p})\:V^{n}_{\mathbf{p},\mathcal{S}}\:e^{-i\mathbf{p}\cdot\mathbf{r}}\bigr],\nonumber
\end{equation}
where $V^{n}_{\mathbf{p},\mathcal{S}}$ is the antineutron eigenstate of spin $\mathcal{S}$ and momentum $\mathbf{p}$, and the creation/annihilation operators satisfy the usual anticommutation relations,
\begin{equation}
\{a^{n}_{\mathcal{S}}(\mathbf{p}),a^{n\dagger}_{\mathcal{S}'}(\mathbf{p}')\}=\{b^{n}_{\mathcal{S}}(\mathbf{p}),b^{n\dagger}_{\mathcal{S}'}(\mathbf{p}')\}=(2\pi)^{3}\delta^{(3)}(\mathbf{p}-\mathbf{p}')\delta_{\mathcal{S}\mathcal{S}'}.\nonumber
\end{equation}
In terms of neutron creation operators acting on the hadronic vacuum, $|\Omega_{h}\rangle_{\mathbf{B}}$, nonrelativistic neutron states read $|\mathbf{p};\mathcal{S}\rangle^{n}_{\mathbf{B}}=\sqrt{2E_{\mathcal{S}}^{n}}\:a^{n\dagger}_{\mathcal{S}}(\mathbf{p})|\Omega_{h}\rangle_{\mathbf{B}}.$\\\\

Lastly,  the  eigenstates of  neutrinos, which are ordinary free massless spinors, are
\[
U^{\nu}_{\mathcal{S}}=\sqrt{(E_{q}+q_{z})/2}\left[ \begin{array}{c}\mathcal{S}\frac{\sqrt{E_{q}^{2}-q_{z}^{2}}}{q_{x}+iq_{y}}\\\\\mathcal{S}\frac{\sqrt{E_{q}-q_{z}}}{\sqrt{E_{q}+q_{z}}}\\\\\frac{q_{z}-E_{q}}{q_{x}+iq_{y}}\\\\1\end{array} \right], \textrm{ for }\mathcal{S}=\pm1,\:\:E_{q}=q.
\]
The neutrino field $\Psi_{\nu}(\mathbf{r})$ reads 
\begin{equation}
\Psi_{\nu}(\mathbf{r})=\sum_{\mathcal{S}=\pm1}\int\frac{\textrm{d}^{3}q}{(2\pi)^{3}}\frac{1}{\sqrt{2E_{q}}}\bigl[a^{\nu}_{\mathcal{S}}(\mathbf{q})\:U^{\nu}_{\mathbf{q},\mathcal{S}}\:e^{i\mathbf{q}\cdot\mathbf{r}}+b^{\nu\dagger}_{\mathcal{S}}(\mathbf{q})\:V^{\nu}_{\mathbf{q},\mathcal{S}}\:e^{-i\mathbf{q}\cdot\mathbf{r}}\bigr],\nonumber
\end{equation}
where $V^{\nu}_{\mathbf{q},\mathcal{S}}$ is the antineutrino eigenstate of spin $\mathcal{S}$ and momentum $\mathbf{q}$, and the creation/annihilation operators satisfy the anticommutation relations,
\begin{equation}
\{a^{\nu}_{\mathcal{S}}(\mathbf{q}),a^{\nu\dagger}_{\mathcal{S}'}(\mathbf{q}')\}=\{b^{\nu}_{\mathcal{S}}(\mathbf{q}),b^{n\dagger}_{\mathcal{S}'}(\mathbf{q}')\}=(2\pi)^{3}\delta^{(3)}(\mathbf{q}-\mathbf{q}')\delta_{\mathcal{S}\mathcal{S}'}.\nonumber
\end{equation}
In terms of neutrino creation operators acting on the leptonic vacuum, $|\Omega_{l}\rangle_{\mathbf{B}}$, neutrino states read $|\mathbf{q};\mathcal{S}\rangle^{\nu}_{\mathbf{B}}=\sqrt{2E_{q}}\:a^{\nu\dagger}_{\mathcal{S}}(\mathbf{q})|\Omega_{l}\rangle_{\mathbf{B}}.$

\section{The electromagnetic field}\label{AppA4}

The free electromagnetic field Hamiltonian which enters Sec.\ref{spinrelax} contains both  the time-derivative and the rotational of the quantum EM field $\mathbf{A}(\mathbf{R})$, which stand for the electric and magnetic quantum fields respectively. In the Coulomb gauge, compatible with the gauge chosen for $\tilde{\mathbf{A}}$, those fields read, in terms of creation and annihilation photon operators \cite{Milonnibook}, 
\begin{eqnarray}
\partial_{t}\mathbf{A}(\mathbf{R})&=&i\sum_{\mathbf{k},s=\pm1}\sqrt{k/2L^{3}}
[\mathbf{\epsilon}^{\mathbf{k}}_{s}a^{\gamma}_{s}(\mathbf{k})e^{i\mathbf{k}\cdot\mathbf{R}}-\mathbf{\epsilon}^{\mathbf{k}*}_{s}a^{\gamma\dagger}_{s}(\mathbf{k})e^{-i\mathbf{k}\cdot\mathbf{R}}],\nonumber\\\nabla\times\mathbf{A}(\mathbf{R})&=&i\sum_{\mathbf{k},s=\pm1}\frac{\mathbf{k}}{\sqrt{2kL^{3}}}\times
[\mathbf{\epsilon}^{\mathbf{k}}_{s}a^{\gamma}_{s}(\mathbf{k})e^{i\mathbf{k}\cdot\mathbf{R}}-\mathbf{\epsilon}^{\mathbf{k}*}_{s}a^{\gamma\dagger}_{s}(\mathbf{k})e^{-i\mathbf{k}\cdot\mathbf{R}}],\nonumber
\end{eqnarray}
where $\mathbf{\epsilon}^{\mathbf{k}}_{s}$ and $a^{\gamma(\dagger)}_{s}(\mathbf{k})$ are the polarisation vector and annihilation, creation $(\dagger)$ operators of photons of momentum $\mathbf{k}$ and polarisation $s$, respectively. The operators satisfy the usual commutaion relations for bosons, $[a^{\gamma}_{s}(\mathbf{k}),a^{\gamma\dagger}_{s'}(\mathbf{k}')]=(2\pi)^{3}\delta^{(3)}(\mathbf{k}-\mathbf{k}')\delta_{ss'}$.

In terms of photon creation operators acting on the EM vacuum, $|\Omega_{\gamma}\rangle$, one-photon states read $|\mathbf{k};s\rangle^{\gamma}=\sqrt{2k}\:a^{\gamma\dagger}_{s}(\mathbf{k})|\Omega_{\gamma}\rangle$; and the free EM Hamiltonian is 
\begin{equation}
H_{EM}=\sum_{\mathbf{k},s=\pm1}k\:[1/2+a^{\gamma\dagger}_{s}(\mathbf{k})a^{\gamma}_{s}(\mathbf{k})].
\end{equation}

\section{Spin expectation values and lepton kinetic momentum operators}\label{AppB}

We use the expressions for the spinor eigenstates of the electroweak particles to compute the spin expectation values which appear in Eq.(\ref{spins}) as well as the lepton kinetic operators, $\mathbf{K}_{e^{+}}$ and $\mathbf{Q}_{\nu}$.

\subsection{Spin expectation values}\label{spines}

Using the equations for the spinor eigenstates in Appendix \ref{AppA}, we write down the expressions for the expectation values of the spin operators given in Eq.(\ref{spins}),
\begin{equation}
\langle\delta\mathbf{S}_{e^{+}}^{\mathcal{N}}\rangle\equiv\frac{\hat{\mathbf{z}}}{2E^{e}_{\mathcal{N}}}V_{\mathcal{N},+}^{e\dagger}S_{z}^{e^{+}}V_{\mathcal{N},+}^{e}+\frac{\hat{\mathbf{z}}}{2E^{e}_{\mathcal{N}-1}}V_{\mathcal{N}-1,-}^{e\dagger}S_{z}^{e^{+}}V_{\mathcal{N}-1,-}^{e}=\frac{e\mathbf{B}}{2(E^{e}_{\mathcal{N}})^{2}}I_{\mathcal{N}}(\xi_{e^{+}})I_{\mathcal{N}}(\xi^{'}_{e^{+}}),\nonumber
\end{equation}
\begin{equation}
\langle\delta\mathbf{S}_{p}\rangle\equiv\frac{\hat{\mathbf{z}}}{2E^{p}_{+}}U_{0,+}^{p\dagger}S_{z}^{p}\:U_{0,+}^{p}+\frac{\hat{\mathbf{z}}}{2E^{p}_{-}}U_{0,-}^{p\dagger}S_{z}^{p}\:U_{0,-}^{p}=\frac{\mu_{N}\mathbf{B}}{2m_{p}}[I_{0}(\xi_{p})I_{0}(\xi^{'}_{p})+I_{1}(\xi_{p})I_{1}(\xi^{'}_{p})],\nonumber
\end{equation}
\begin{equation}
\langle\delta\mathbf{S}_{n}\rangle\equiv\frac{\hat{\mathbf{z}}}{2E^{n}_{+}}U_{+}^{n\dagger}S_{z}^{n}\:U_{+}^{n}+\frac{\hat{\mathbf{z}}}{2E^{n}_{-}}U_{-}^{n\dagger}S_{z}^{n}\:U_{-}^{n}=\frac{|\kappa_{n}|\mu_{N}\mathbf{B}}{4m_{n}},\nonumber
\end{equation}
where $S_{z}^{e^{+}}$,  $S_{z}^{p}$ and  $S_{z}^{n}$ are the components of the spin operators along $\mathbf{B}$, for positrons, protons and neutrons, respectively.

\subsection{Kinetic momentum operators of leptons}\label{KQ}

In this subsection we write down the expressions of the kinetic momentum operators of positrons and neutrinos, $\mathbf{K}_{e^{+}}$ and $\mathbf{Q}_{\nu}$, in terms of their creation and annihilation operators. 

As for the neutrino momentum, it is,
\begin{equation}
\mathbf{Q}_{\nu}=\int\textrm{d}^{3}r\:\Psi^{\dagger}_{\nu}(\mathbf{r})(-i\mathbf{\nabla})\Psi_{\nu}(\mathbf{r})|_{\textrm{neutrinos}}=\int\frac{\textrm{d}^{3}q}{(2\pi)^{3}}\mathbf{q}\sum_{\mathcal{S}=\pm1}a^{\nu\dagger}_{\mathcal{S}}(\mathbf{q})a^{\nu}_{\mathcal{S}}(\mathbf{q}).\label{Qn}
\end{equation}

As for the kinetic momentum of positrons in a magnetic field $\mathbf{B}$, we make use of the formulas developed in Appendix \ref{posit} in the gauge $\tilde{\mathbf{A}}=-Br_{y}\hat{\mathbf{x}}$.
\begin{align}
\mathbf{K}_{e^{+}}&=\int\textrm{d}^{3}r\:[\Psi^{\dagger}_{e}(\mathbf{r})(-i\mathbf{\nabla})\Psi_{e}(\mathbf{r})+e\tilde{\mathbf{A}}]|_{\textrm{positrons}}=\hat{\mathbf{z}}\int\frac{\textrm{d}k_{x}\textrm{d}k_{z}}{(2\pi)^{2}}k_{z}\sum_{\mathcal{S}=\pm1}\sum_{\mathcal{N}=0}b^{e\dagger}_{\mathcal{N},\mathcal{S}}(k_{x},k_{z})b^{e}_{\mathcal{N},\mathcal{S}}(k_{x},k_{z})\label{Ke}\\
&+\int\frac{\textrm{d}k_{x}\textrm{d}k_{z}}{(2\pi)^{2}}k_{z}\sum_{\mathcal{N}=1}\Bigl[2\sqrt{E_{\mathcal{N}}^{e}E_{\mathcal{N}-1}^{e}(E_{\mathcal{N}}^{e}+m_{e})(E_{\mathcal{N}-1}^{e}+m_{e})}\Bigr]^{-1}\Bigl\{\frac{\sqrt{eB(\mathcal{N}-1)}}{2\sqrt{2}}\mathcal{F}_{\mathcal{N}}(eB)\Bigl[(\hat{\mathbf{x}}+i\hat{\mathbf{y}})b^{e\dagger}_{(\mathcal{N},+)}b^{e}_{(\mathcal{N}-1,+)}\nonumber\\
&+(\hat{\mathbf{x}}-i\hat{\mathbf{y}})b^{e\dagger}_{(\mathcal{N}-1,+)}b^{e}_{(\mathcal{N},+)}\Bigr]+\frac{\sqrt{eB\mathcal{N}}}{2\sqrt{2}}[\mathcal{F}_{\mathcal{N}}(eB)-4eB]\Bigl[(\hat{\mathbf{x}}+i\hat{\mathbf{y}})b^{e\dagger}_{(\mathcal{N},-)}b^{e}_{(\mathcal{N}-1,-)}+(\hat{\mathbf{x}}-i\hat{\mathbf{y}})b^{e\dagger}_{(\mathcal{N}-1,-)}b^{e}_{(\mathcal{N},-)}\Bigr]\nonumber\\
&-k_{z}eB\Bigl[(\hat{\mathbf{x}}+i\hat{\mathbf{y}})b^{e\dagger}_{(\mathcal{N},+)}b^{e}_{(\mathcal{N}-1,-)}+(\hat{\mathbf{x}}-i\hat{\mathbf{y}})b^{e\dagger}_{(\mathcal{N}-1,-)}b^{e}_{(\mathcal{N},+)}\Bigr]\Bigr\},\nonumber
\end{align}
where $\mathcal{F}_{\mathcal{N}}(eB)\equiv(E_{\mathcal{N}}^{e}+m_{e}+k_{z})(E_{\mathcal{N}-1}^{e}+m_{e}+k_{z})+(E_{\mathcal{N}}^{e}+m_{e}-k_{z})(E_{\mathcal{N}-1}^{e}+m_{e}-k_{z})+4\mathcal{N}eB$, and the dependence of the creation and annihilation positron operators on $k_{x}$ and $k_{z}$ has been omitted for brevity.

\section{Quadratic fluctuations of leptonic and hadronic currents}\label{AppC}

The hadronic and leptonic currents flanking the energy denominator of the integrands in Eqs.(\ref{equationmother}) and (\ref{equationmother2}) give rise to quadratic vacuum fluctuations for the case of leptons, and quadratic hadron fluctuations upon the single proton states $|\phi^{p}_{ext}\rangle_{\mathbf{B}_{0}}\equiv\bigl[\bigl|0,1;\langle\mathbf{P}_{N}\rangle\bigl\rangle+|0,-1;\langle\mathbf{P}_{N}\rangle\bigl\rangle\bigr]^{p}_{\mathbf{B}_{0}}/2L\sqrt{E_{0}^{p}}$ in  Eq.(\ref{equationmother}), and $|\phi^{p'}_{ext}(t)\rangle_{\mathbf{B}_{0}}\equiv\bigl[(2-e^{-\Gamma_{p}t})^{1/2}|0,1;\langle\mathbf{P}_{N}\rangle/\sqrt{E^{p}_{+}}+e^{-\Gamma_{p}t/2}|0,-1;\langle\mathbf{P}_{N}\rangle/\sqrt{E^{p}_{-}}\bigr]^{p}_{\mathbf{B}_{0}}/2L$ in Eq.(\ref{equationmother2}), respectively. In either case, we write the equations of these fluctuations as integrals over the momenta of the virtual particles.

%and upon the state  $|\phi^{p}_{ext}(t)\rangle=\bigl[(2-e^{-\Gamma_{p}t})^{1/2}|0,1;\langle\mathbf{P}_{N}\rangle/\sqrt{E^{p}_{+}}+e^{-\Gamma_{p}t/2}|0,-1;\langle\mathbf{P}_{N}\rangle/\sqrt{E^{p}_{-}}\bigr]^{p}_{\mathbf{B}_{0}}/2L$ of a partially polarized proton in Eq.(6). Here $L$ is a quantization length to be taken to infinity, $L\rightarrow\infty$. In either case, we write the equations of these fluctuations as integrals over the momenta of the virtual particles.

\subsection{Leptonic vacuum fluctuations}

The equation for the leptonic current fluctuations in the leptonic vacuum state, $|\Omega_{l}\rangle_{\mathbf{B}_{0}}$, is  

% ALL THE LEVI CIVITA TENSORS IN EN QUADRATIC LEPTONIC AND HADRONIC FLUCTUATIONS HAVE BEEN ALREADY CHANGED
\begin{align}
&_{\mathbf{B}_{0}}\langle\Omega_{l}|J^{l}_{\mu}(\mathbf{r})J^{l\dagger}_{\rho}(\mathbf{r}')|\Omega_{l}\rangle_{\mathbf{B}_{0}}=\int\frac{\textrm{d}^{3}q}{(2\pi)^{3}}\frac{\textrm{d}k_{1}\textrm{d}k_{3}}{(2\pi)^{2}}\frac{e^{i(\mathbf{r}-\mathbf{r}')\cdot(\mathbf{q}+\mathbf{k}')}e^{-i(r_{2}-r_{2}')k_{2}}}{2E_{q}2E^{e}_{\mathcal{N}}}\sum_{\mathcal{N}=0}^{\infty}\sum_{s,r=\pm1}\textrm{Tr}\{\bar{V}_{\mathcal{N},s}^{e^{+}}[\gamma_{\mu}(\mathbb{I}-\gamma_{5})]U^{\nu}_{r}\bar{U}^{\nu}_{r}[\gamma_{\rho}(\mathbb{I}-\gamma_{5})]V_{\mathcal{N},s}^{e^{+}}\}\nonumber\\
&=\int\frac{\textrm{d}^{3}q}{(2\pi)^{3}q}\frac{\textrm{d}k_{x}\textrm{d}k_{z}}{(2\pi)^{2}}e^{i(\mathbf{r}-\mathbf{r}')\cdot(\mathbf{q}+\mathbf{k}')}e^{-i(r_{2}-r_{2}')k_{2}}\Big\{I_{0}(\xi_{e^{+}})I_{0}(\xi'_{e^{+}})(1+k_{3}/E_{0}^{e})\Big[q(2\delta_{\mu0}\delta_{\rho0}-g_{\mu\rho}-\delta_{\mu0}\delta_{\rho3}-\delta_{\mu3}\delta_{\rho0}+i\epsilon_{0\mu\rho3})\nonumber\\
&+ q^{i}(\delta_{\mu0}\delta_{\rho i}+\delta_{\mu i}\delta_{\rho0}+i\epsilon_{0\mu\rho i}-\delta_{\mu i}\delta_{\rho3}-\delta_{\mu3}\delta_{\rho i}-g_{\mu\rho}\delta_{i3}-i\epsilon_{\mu\rho i3})\Big]+I_{0}(\xi_{e^{+}})I_{0}(\xi'_{e^{+}})(1-k_{3}/E_{1}^{e})\Big[q(2\delta_{\mu0}\delta_{\rho0}-g_{\mu\rho}\nonumber\\
&+\delta_{\mu0}\delta_{\rho3}+\delta_{\mu3}\delta_{\rho0}-i\epsilon_{0\mu\rho3})+ q^{i}(\delta_{\mu0}\delta_{\rho i}+\delta_{\mu i}\delta_{\rho0}+i\epsilon_{0\mu\rho i}+\delta_{\mu i}\delta_{\rho3}+\delta_{\mu3}\delta_{\rho i}+g_{\mu\rho}\delta_{i3}+i\epsilon_{\mu\rho i3})\Big]+\sum_{\mathcal{N}=1}^{\infty}I_{\mathcal{N}}(\xi_{e^{+}})I_{\mathcal{N}}(\xi'_{e^{+}})\nonumber\\
&\times[k_{3}eB_{0} /(E_{\mathcal{N}}^{e})^{3}]\Big[q(2\delta_{\mu0}\delta_{\rho0}-g_{\mu\rho}+\delta_{\mu0}\delta_{\rho3}+\delta_{\mu3}\delta_{\rho0}-i\epsilon_{0\mu\rho3})+ q^{i}(\delta_{\mu0}\delta_{\rho i}+\delta_{\mu i}\delta_{\rho0}+i\epsilon_{0\mu\rho i}+\delta_{\mu i}\delta_{\rho3}+\delta_{\mu3}\delta_{\rho i}+g_{\mu\rho}\delta_{i3}\nonumber\\
&+i\epsilon_{\mu\rho i3})\Big]+2I_{\mathcal{N}}(\xi_{e^{+}})I_{\mathcal{N}}(\xi'_{e^{+}})\Big[q(2\delta_{\mu0}\delta_{\rho0}-g_{\mu\rho})-(k_{3}q/E_{\mathcal{N}}^{e})(\delta_{\mu0}\delta_{\rho3}+\delta_{\mu3}\delta_{\rho0}-i\epsilon_{0\mu\rho3})+ q^{i}(\delta_{\mu0}\delta_{\rho i}+\delta_{\mu i}\delta_{\rho0}+i\epsilon_{0\mu\rho i})\nonumber\\
&-(k_{3}q^{i}/E_{\mathcal{N}}^{e})(\delta_{\mu i}\delta_{\rho3}+\delta_{\mu3}\delta_{\rho i}+g_{\mu\rho}\delta_{i3}+i\epsilon_{\mu\rho i3})\Big] - \sum_{\mathcal{N}=1}^{\infty}(\sqrt{2\mathcal{N}eB_{0}}/E_{\mathcal{N}}^{e})[I_{\mathcal{N}}(\xi_{e^{+}})I_{\mathcal{N}-1}(\xi'_{e^{+}})+I_{\mathcal{N}-1}(\xi_{e^{+}})I_{\mathcal{N}}(\xi'_{e^{+}}]\nonumber\\
&\times\Big[q(\delta_{\mu1}\delta_{\rho0}+\delta_{\mu0}\delta_{\rho1}+i\epsilon_{1\mu\rho0})+q^{i}(\delta_{\mu1}\delta_{\rho i}+\delta_{\mu i}\delta_{\rho1}+g_{\mu\rho}\delta_{i1}+i\epsilon_{1\mu\rho0})\Big]
+i \sum_{\mathcal{N}=1}^{\infty}(\sqrt{2\mathcal{N}eB_{0}}/E_{\mathcal{N}}^{e})\nonumber\\
&\times[I_{\mathcal{N}}(\xi_{e^{+}})I_{\mathcal{N}-1}(\xi'_{e^{+}})-I_{\mathcal{N}-1}(\xi_{e^{+}})I_{\mathcal{N}}(\xi'_{e^{+}}]\Big[q(\delta_{\mu2}\delta_{\rho0}+\delta_{\mu0}\delta_{\rho2}+i\epsilon_{2\mu\rho0})+q^{i}(\delta_{\mu2}\delta_{\rho i}+\delta_{\mu i}\delta_{\rho2}+g_{\mu\rho}\delta_{i2}+i\epsilon_{2\mu\rho0})\Big]\Big\},\label{SM3}
\end{align}
where greek indices take the values $0,1,2,3$, whereas latin indices take $1,2,3$. The index $0$ corresponds to the time component, while $1,2,3$ refer to the spatial cartesian components $x,y,z$, respectively. In Eq.(\ref{SM3}) the factors proportional to $eB_{0}/(E_{\mathcal{N}}^{e})^{2}$ result from the approximation $E_{\mathcal{N}}^{e}/E_{\mathcal{N}+1}^{e}\simeq1-eB_{0}/(E_{\mathcal{N}}^{e})^{2}$. It is only the terms proportional to $(1+k_{3}/E_{0}^{e})$ in the second and third rows that correspond to fluctuations of positrons in the lowest Landau level.

\subsection{Hadronic current fluctuations}

As for the hadronic current fluctuations which enter Eq.(\ref{equationmother2}) for the case that the proton state is $|\phi_{ext}^{p'}(t)\rangle_{\mathbf{B}_{0}}$, $t> t_{0}$, $t_{0}\ll\Gamma_{p}^{-1}$,
%Leyendo de la ecuacion H.6 p.10 dated 10/05=2018, and then passing to Eq.(H.15) on p.5 dated 25/05/2018 

\begin{align}
&\:_{\mathbf{B}_{0}}\langle\phi_{ext}^{p'}(t)|J^{h}_{\mu}(\mathbf{r})J^{h\dagger}_{\rho}(\mathbf{r}')|\phi_{ext}^{p'}(t)\rangle_{\mathbf{B}_{0}}=\int\frac{\textrm{d}^{3}p}{(2\pi)^{3}}\sum_{s,r=\pm1}\frac{e^{-\Gamma_{p}t}}{2L^{2}}\frac{e^{i(\mathbf{r}-\mathbf{r}')\cdot\mathbf{p}}}{2E_{r}^{n}2E^{p}_{s}}\textrm{Tr}\{\bar{U}_{0,s}^{p}[\gamma_{\mu}(\mathbb{I}-g_{A}\gamma_{5})]U^{n}_{r}\bar{U}^{n}_{r}[\gamma_{\rho}(\mathbb{I}-g_{A}\gamma_{5})]U_{0,s}^{p}\}\nonumber\\
&+\int\frac{\textrm{d}^{3}p}{(2\pi)^{3}}\sum_{r=\pm1}\frac{1-e^{-\Gamma_{p}t}}{L^{2}}\frac{e^{i(\mathbf{r}-\mathbf{r}')\cdot\mathbf{p}}}{2E_{r}^{n}2E^{p}_{+}}\textrm{Tr}\{\bar{U}_{0,+}^{p}[\gamma_{\mu}(\mathbb{I}-g_{A}\gamma_{5})]U^{n}_{r}\bar{U}^{n}_{r}[\gamma_{\rho}(\mathbb{I}-g_{A}\gamma_{5})]U_{0,+}^{p}\}\nonumber\\
&=\frac{e^{-\Gamma_{p}t}}{L^{2}}\int\frac{\textrm{d}^{3}p}{(2\pi)^{3}}e^{i(\mathbf{r}-\mathbf{r}')\cdot\mathbf{p}}I_{0}(r_{2})I_{0}(r'_{2})\Big[\delta_{\mu0}\delta_{\rho0}+g_{A}^{2}(\delta_{\mu0}\delta_{\rho0}-g_{\mu\rho})+\frac{{p}^{i}}{2m_{n}}[(1+g_{A}^{2})(\delta_{\mu0}\delta_{\rho i}+\delta_{\mu i}\delta_{\rho0})-2g_{A}i\epsilon_{0\mu i\rho}]\Big]\nonumber\\
& + \frac{1-e^{-\Gamma_{p}t}}{2L^{2}}\int\frac{\textrm{d}^{3}p}{(2\pi)^{3}}e^{i(\mathbf{r}-\mathbf{r}')\cdot\mathbf{p}}I_{0}(r_{2})I_{0}(r'_{2})\Big[(1-g_{A}^{2})g_{\mu\rho}+(1+g_{A}^{2})(2\delta_{\mu0}\delta_{\rho0}-g_{\mu\rho}) + 2g_{A}(\delta_{\mu3}\delta_{\rho0}+\delta_{\mu0}\delta_{\rho3})\nonumber\\
&-2g_{A}^{2}i\epsilon_{0\mu\rho3}+\frac{{p}^{i}}{m_{n}}[(1+g_{A}^{2})(\delta_{\mu0}\delta_{\rho i}+\delta_{\mu i}\delta_{\rho0}+i\epsilon_{\mu\rho i3})+2g_{A}(\delta_{\mu3}\delta_{\rho i}+\delta_{\mu i}\delta_{\rho3}+\delta_{i3}g_{\mu\rho}+i\epsilon_{0\mu\rho j})]\Big]+\mathcal{O}(eB_{0}/m_{p}^{2}),\label{SM4}
\end{align}
where terms of the order of $eB_{0}/m_{p}^{2}$ smaller have been discarded since they are irrelevant for the computation of the Casimir momentum of Eqs.(\ref{f2a}) and (\ref{f2b}).

For the hadron current fluctuations in Eq.(\ref{equationmother}), the external proton state is the stationary state $|\phi^{p}_{ext}\rangle_{\mathbf{B}_{0}}$, and the current fluctuations read, up to terms of order $eB_{0}/m_{p}^{2}$,

\begin{align}
&_{\mathbf{B}_{0}}\langle\phi^{p}_{ext}|J^{h}_{\mu}(\mathbf{r})J^{h\dagger}_{\rho}(\mathbf{r}')|\phi^{p}_{ext}\rangle_{\mathbf{B}_{0}}=\int\frac{\textrm{d}^{3}p}{(2\pi)^{3}}\sum_{s,r=\pm1}\frac{e^{i(\mathbf{r}-\mathbf{r}')\cdot\mathbf{p}}}{2L^{2}2E_{r}^{n}2E^{p}_{s}}\textrm{Tr}\{\bar{U}_{0,s}^{p}[\gamma_{\mu}(\mathbb{I}-g_{A}\gamma_{5})]U^{n}_{r}\bar{U}^{n}_{r}[\gamma_{\rho}(\mathbb{I}-g_{A}\gamma_{5})]U_{0,s}^{p}\}\nonumber\\
&=\int\frac{\textrm{d}^{3}p}{(2\pi)^{3}}\frac{e^{i(\mathbf{r}-\mathbf{r}')\cdot\mathbf{p}}}{4L{2}}\Big\{4I_{0}(r_{2})I_{0}(r'_{2})\Big[\delta_{\mu0}\delta_{\rho0}+g_{A}^{2}(\delta_{\mu0}\delta_{\rho0}-g_{\mu\rho})+\frac{{p}^{i}}{2m_{n}}[(1+g_{A}^{2})(\delta_{\mu0}\delta_{\rho i}+\delta_{\mu i}\delta_{\rho0})-2g_{A}i\epsilon_{0\mu i\rho}]\Big]\nonumber\\
&-\frac{eB_{0}}{2m_{p}^{2}}[I_{0}(r_{2})I_{0}(r'_{2})+I_{1}(r_{2})I_{1}(r'_{2})]\nonumber\\
&\times\Big[(1-g_{A}^{2})g_{\mu\rho}-2g_{A}(\delta_{\mu3}\delta_{\rho0}+\delta_{\mu0}\delta_{\rho3})+(1+g_{A}^{2})i\epsilon_{0\mu\rho3}+\frac{p^{i}}{m_{n}}[ (1+g_{A}^{2})i\epsilon_{\mu\rho i3} - 2g_{A}(\delta_{\mu3}\delta_{\rho i}+\delta_{\mu i}\delta_{\rho3}+ \delta_{i3}g_{\mu\rho})]\Big]\nonumber\\ 
&+ \frac{eB_{0}}{2m_{p}^{2}}[I_{1}(r_{2})I_{1}(r'_{2})-I_{0}(r_{2})I_{0}(r'_{2})]\Big[(1+g_{A}^{2})[2\delta_{\mu0}\delta_{\rho0}-g_{\mu\rho}+\frac{p^{i}}{m_{n}}((\delta_{\mu0}\delta_{\rho i}+\delta_{\mu i}\delta_{\rho0})]-(1+g_{A}^{2})i\epsilon_{03\mu\rho}\Big]\nonumber\\
&- \frac{\sqrt{eB_{0}}}{\sqrt{2}m_{p}}[I_{0}(r_{2})I_{1}(r'_{2})+I_{1}(r_{2})I_{0}(r'_{2})]\Big[(1+g_{A}^{2})(\delta_{\mu2}\delta_{\rho0}+\delta_{\mu0}\delta_{\rho2})-2g_{A}i\epsilon_{02\mu\rho}+\frac{p^{i}}{m_{n}}[(1+g_{A}^{2})(\delta_{\mu2}\delta_{\rho i}+\delta_{\mu i}\delta_{\rho2}\nonumber\\
&+ \delta_{i2}g_{\mu\rho}) + 2g_{A}i\epsilon_{i2\mu\rho}]\Big]+4i\frac{\sqrt{eB_{0}}}{\sqrt{2}m_{p}}[I_{0}(r_{2})I_{1}(r'_{2})+I_{1}(r_{2})I_{0}(r'_{2})](1-g_{A}^{2})(\delta_{\mu0}\delta_{\rho1}-\delta_{\mu1}\delta_{\rho0})+\frac{|g_{n}|\mu_{N}B_{0}}{2m_{p}^{2}}I_{0}(r_{2})I_{0}(r'_{2})\nonumber\\
&\times\Big[(1+g_{A})^{2}(2\delta_{\mu0}\delta_{\rho0}-\delta_{\mu0}\delta_{\rho3}-\delta_{\mu3}\delta_{\rho0})-2g_{A}(1+g_{A})(g_{\mu\rho}+i\epsilon_{03\mu\rho}) + \frac{p_{3}}{2m_{n}}[(1-g_{A}^{2})g_{\mu\rho} + (1+3g_{A}^{2})i \epsilon_{03\mu\rho}\nonumber\\
&+4g_{A}(2\delta_{\mu0}\delta_{\rho0}+\delta_{\mu0}\delta_{\rho3}+\delta_{\mu3}\delta_{\rho0}-g_{\mu\rho})]+ \frac{p_{1}}{2m_{n}}[(g_{A}^{2}-1)(\delta_{\mu2}\delta_{\rho0}-\delta_{\mu0}\delta_{\rho2} - i\epsilon_{01\mu\rho})+(1+g_{A})^{2}(\delta_{\mu0}\delta_{\rho1}+\delta_{\mu1}\delta_{\rho0}\nonumber\\
&+i\epsilon_{01\mu\rho})]+\frac{p_{2}}{2m_{n}}[(g_{A}^{2}-1)(\delta_{\mu1}\delta_{\rho0}-\delta_{\mu0}\delta_{\rho1}-i\epsilon_{02\mu\rho})+(1+g_{A})^{2}(\delta_{\mu0}\delta_{\rho2}+\delta_{\mu2}\delta_{\rho0}+i\epsilon_{02\mu\rho})]\Big]\Big\}.\label{SM5}\nonumber
\end{align}

\section{Kinetic energy terms in $\mathcal{E}_{\mathcal{N},m}$ and form factors}\label{AppD}

In this Appendix we derive the kinetic terms which enter the factor $\mathcal{E}_{\mathcal{N},m}$ in the denominator of the integrands of Eqs.(\ref{f1a}) and (\ref{f2a}), as well as the form factors $\langle\tilde{\phi}^{0}_{int}|\tilde{\phi}_{int}^{m}\rangle$ which appear in those equations. %To this aim, we write the total kinetic energy of the quarks within the nucleons in terms of Jacobi variables in the nonrelativistic limit.

\subsection{Kinetic energy terms in $\mathcal{E}_{\mathcal{N},m}$}\label{AppD1}
We follow an approach analogous to that in chapter 3 of Ref.\cite{Milonnibook}, P.W. Milonni, and references therein, which is applied there to the computation of the Lamb-shift. Let us consider the integrand of Eq.(\ref{equationmother}), and let us evaluate the energy denominator there for certain intermediate state of positrons, neutrinos and neutrons $|I_{e,\nu,n}\rangle_{\mathbf{B}_{0}}= |\mathcal{N},\mathcal{S}_{e^{+}};k_{x},k_{z}\rangle_{\mathbf{B}_{0}}^{e}\otimes|\mathbf{q},\mathcal{S}_{\nu}\rangle^{\nu}\otimes|\phi_{int}^{m}\rangle\otimes|\mathbf{p},\mathcal{S}\rangle^{n}_{\mathbf{B}_{0}}$,
\begin{equation}
_{\mathbf{B}_{0}}\langle I_{e,\nu,n}|[H_{int}^{N}+H_{D}^{n}+H_{D}^{e}+H_{D}^{\nu}-E_{0}^{p}]|I_{e,\nu,n}\rangle_{\mathbf{B}_{0}}=\langle\phi_{int}^{m}|H^{N}_{int}|\phi_{int}^{m}\rangle+E_{\mathcal{S}}^{n}+E_{\mathcal{N}}^{e}+E_{q}-E_{0}^{p},\label{SM7}
\end{equation}
where the expressions for $\langle\phi_{int}^{m}|H^{N}_{int}|\phi_{int}^{m}\rangle$, $E_{\mathcal{S}}^{n}$, $E_{\mathcal{N}}^{e}$, $E_{q}$ and $E_{0}^{p}$ are given in the Sec.\ref{Approach}. Let us substitute the expression in Eq.(\ref{SM7}) into the integrand of Eq.(\ref{f1a}), including  the spatial wavefunction of the  $\mathcal{N}^{th}$ positron Landau level, $e^{-\tilde{\xi}^{2}_{e^{+}}/2}H_{\mathcal{N}}(\tilde{\xi}_{e^{+}})$,  and the integrals over the coordinates $x$, $z$, $p_{x}$ and $p_{z}$ of the nucleon centre of mass,
\begin{align}
&\int\frac{\textrm{d}x\textrm{d}z\textrm{d}p_{x}\textrm{d}p_{z}}{(2\pi)^{2}}\langle\phi^{0}_{int}|e^{i\tilde{\mathbf{r}}\cdot(\mathbf{p}+\mathbf{k}+\mathbf{q})-ik_{y}\tilde{r}_{y}}e^{-\tilde{\xi}^{2}_{e^{+}}/2}H_{\mathcal{N}}(\tilde{\xi}_{e^{+}})\frac{1}{E^{e}_{\mathcal{N}}+E_{q}+E^{n}_{\mathcal{S}}(\mathbf{p})-E_{0}^{p}+H_{int}^{N}(\mathbf{p}_{\rho};\mathbf{p}_{\lambda})}|\phi_{int}^{m}\rangle\nonumber\\
&\times\hat{\mathbf{z}}\cdot(\mathbf{k}+\mathbf{q})\langle\phi_{int}^{m}|\frac{1}{E^{e}_{\mathcal{N}}+E_{q}+E^{n}_{\mathcal{S}}(\mathbf{p})-E_{0}^{p}+H_{int}^{N}(\mathbf{p}_{\rho};\mathbf{p}_{\lambda})} e^{i\tilde{\mathbf{r}}'\cdot(\mathbf{p}+\mathbf{k}+\mathbf{q})-ik_{y}\tilde{r}_{y}'}e^{-\tilde{\xi}^{'2}_{e^{+}}/2}H_{\mathcal{N}}(\tilde{\xi}'_{e^{+}})|\phi_{int}^{0}\rangle,\label{SM8}
\end{align}
where $\tilde{\xi}_{e^{+}}=\sqrt{eB}\tilde{r}_{y}+k_{x}/\sqrt{eB}$, and we have rewritten the coordinate of the active quark in terms of the nucleon centre of mass $\mathbf{R}$ and the Jacobi coordinate $\mathbf{r}_{\lambda}$, $\tilde{\mathbf{r}}=\mathbf{R}+(2/3)\mathbf{r}_{\lambda}$, and the internal nucleon Hamiltonian of Eq.(\ref{Hint}) as an explicit function of $\mathbf{p}_{\rho}$ and $\mathbf{p}_{\lambda}$. The exponential factors flanking the denominator come from the hadronic and leptonic currents on either side of the integrand in Eq.(\ref{equationmother}). Bearing in mind that the relevant contribution of positrons to Eq.(\ref{f1a}) comes from relativistic positrons, the following approximation can be used for Landau levels with $\mathcal{N}\gg1$,
\begin{equation}
e^{-\tilde{\xi}^{2}_{e^{+}}/2}H_{\mathcal{N}}(\tilde{\xi}_{e^{+}})\simeq e^{-\xi^{2}_{e^{+}}/2}H_{\mathcal{N}}(\xi_{e^{+}})\cos{(2r^{y}_{\lambda}\sqrt{2\mathcal{N}eB}/3)}.\nonumber
\end{equation}
Substituting the above expression into Eq.(\ref{SM8}), we arrive at
\begin{align}
&\int\frac{\textrm{d}x\textrm{d}z\textrm{d}p_{x}\textrm{d}p_{z}}{(2\pi)^{2}}\langle\phi^{0}_{int}|e^{i\mathbf{R}\cdot(\mathbf{p}+\mathbf{k}+\mathbf{q})-ik_{y}R_{y}}e^{i(2/3)[\mathbf{r}_{\lambda}\cdot(\mathbf{k}+\mathbf{q})-k_{y}r_{\lambda}^{y}]}e^{-\xi^{2}_{e^{+}}/2}H_{\mathcal{N}}(\xi_{e^{+}})\cos{(2r^{y}_{\lambda}\sqrt{2\mathcal{N}eB}/3)}\nonumber\\
&\times\frac{1}{E^{e}_{\mathcal{N}}+E_{q}+E^{n}_{\mathcal{S}}(\mathbf{p})-E_{0}^{p}+H_{int}^{N}(\mathbf{p}_{\rho};\mathbf{p}_{\lambda})}|\phi_{int}^{m}\rangle\hat{\mathbf{z}}\cdot(\mathbf{k}+\mathbf{q})\langle\phi_{int}^{m}|\frac{1}{E^{e}_{\mathcal{N}}+E_{q}+E^{n}_{\mathcal{S}}(\mathbf{p})-E_{0}^{p}+H_{int}^{N}(\mathbf{p}_{\rho};\mathbf{p}_{\lambda})}\nonumber\\
&\times e^{-\xi^{'2}_{e^{+}}/2}H_{\mathcal{N}}(\xi'_{e^{+}})\cos{(2r^{y}_{\lambda}\sqrt{2\mathcal{N}eB}/3)} e^{-i\mathbf{R}'\cdot(\mathbf{p}+\mathbf{k}+\mathbf{q})+ik_{y}R'_{y}}e^{-i(2/3)[\mathbf{r}_{\lambda}\cdot(\mathbf{k}+\mathbf{q})-k_{y}r_{\lambda}^{y}]}|\phi_{int}^{0}\rangle.\label{SM9}
\end{align}
The integration in $x$ and $z$ yields the deltas of momentum conservation $(2\pi)^{2}\delta(p_{x}+k_{x}+q_{x})\delta(p_{z}+k_{z}+q_{z})$.  Further, integration in momentum coordinates leads to the following effective replacement in the denominator of Eq.(\ref{SM9}),
 \begin{equation}
 E^{n}_{\mathcal{S}}(p_{x},p_{y},p_{z})\rightarrow E^{n}_{\mathcal{S}}(-k_{x}-q_{x},p_{y},-k_{z}-q_{z}).\label{SM10}
 \end{equation} 
 Next, we apply the canonical commutation relations, $[r_{\lambda}^{i},p_{\lambda}^{j}]=i\delta_{ij}$, to move the $r_{\lambda}^{y}$-dependent exponentials to the middle of Eq.(\ref{SM9}). Making use of the relationship
 \begin{align}
&\langle\phi^{0}_{int}|e^{i(2/3)[\mathbf{r}_{\lambda}\cdot(\mathbf{k}+\mathbf{q})+r_{\lambda}^{y}(\pm\sqrt{2\mathcal{N}eB}r^{y}_{\lambda}-k_{y})]}\frac{1}{E^{e}_{\mathcal{N}}+E_{q}+E^{n}_{\mathcal{S}}-E_{0}^{p}+H_{int}^{N}(\mathbf{p}_{\rho};\mathbf{p}_{\lambda})}|\phi_{int}^{m}\rangle\nonumber\\
&=\frac{1}{E^{e}_{\mathcal{N}}+E_{q}+E^{n}_{\mathcal{S}}-E_{0}^{p}+\langle\phi^{m}_{int}|H_{int}^{N}[\mathbf{p}_{\rho};\mathbf{p}_{\lambda}-2[\mathbf{k}+\mathbf{q}+\hat{\mathbf{y}}(\pm\sqrt{2\mathcal{N}eB}r^{y}_{\lambda}-k_{y})]/3]|\phi_{int}^{m}\rangle}\nonumber\\&\times\langle\phi^{m}_{int}|e^{i(2/3)[\mathbf{r}_{\lambda}\cdot(\mathbf{k}+\mathbf{q})+r_{\lambda}^{y}(\pm\sqrt{2\mathcal{N}eB}r^{y}_{\lambda}-k_{y})]}|\phi_{int}^{0}\rangle,\label{SM11}
\end{align}
and the replacement in Eq.(\ref{SM10}), the energy denominators of Eq.(\ref{SM9}) read, for a nonrelativistic neutron,
\begin{align}
 &E_{n}^{\mathcal{S}}(-x_{x}-q_{x},p_{y},-k_{z}-q_{z})+\langle\phi^{m}_{int}|H_{int}^{N}[\mathbf{p}_{\rho};\mathbf{p}_{\lambda}-2[\mathbf{k}+\mathbf{q}+\hat{\mathbf{y}}(\pm\sqrt{2\mathcal{N}eB}r^{y}_{\lambda}-k_{y})]/3]|\phi_{int}^{m}\rangle+E^{e}_{\mathcal{N}}+E_{q}-E_{0}^{p}\nonumber\\
 &=\langle\phi^{m}_{int}|\Bigl[\frac{|\mathbf{k}+\mathbf{q}+(\pm\sqrt{2\mathcal{N}eB}-k_{y})\hat{\mathbf{y}}|^{2}}{2m_{d}}-\frac{\mathbf{p}_{\lambda}}{m_{d}}\cdot[\mathbf{k}+\mathbf{q}+(\pm\sqrt{2\mathcal{N}eB}-k_{y})\hat{\mathbf{y}}]+V_{\textrm{conf}}^{N}\Bigl]|\phi_{int}^{m}\rangle+E^{e}_{\mathcal{N}}+E_{q}\nonumber\\
 &-(\mathcal{S}g_{n}+1)\mu_{N}B+(m_{n}-m_{p})+\frac{p^{2}_{y}}{2m_{n}}.\label{SM12}
 \end{align}
where we have replaced $H_{int}^{N}$ with its expression in Eq.(\ref{Hint}). Further, we observe in Eqs.(\ref{f1a}) and (\ref{f2a}) that all the nonvanishing integrals in $y$ and $y'$  yield factors of the form 
\begin{equation}
\int_{-\infty}^{\infty}I_{0,1}(y)I_{\mathcal{N}}(\xi_{e^{+}})e^{iy(p_{y}+q_{y})}\textrm{d}y\propto e^{-[k_{x}^{2}+(p_{y}+q_{y})^{2}]/2eB},\nonumber
\end{equation}
which suppress exponentially the remaining integrands on $k_{x}$, $p_{y}$ and $q_{y}$, for $|k_{x}|>\sqrt{eB}$ and $|p_{y}+q_{y}|>\sqrt{eB}$. In particular, for $eB\ll m^{2}_{d},m^{2}_{n}$, this implies that the terms proportional to $k_{x}/m_{d}$ are negligible and that we can replace $p_{y}^{2}/2m_{n}\rightarrow q_{y}^{2}/2m_{n}$ at the same order of approximation in Eq.(\ref{SM12}) . In addition, the terms $\mathbf{p}_{\lambda}\cdot[\mathbf{k}+\mathbf{q}+(\pm\sqrt{2\mathcal{N}eB}-k_{y})\hat{\mathbf{y}}]/m_{d}$ are Doppler shift terms which are negligible in the nonrelativistic limit, and so are the spin-dependent magnetic terms. As a result, the energy denominators of Eq.(\ref{SM11}) can be approximated by
 \begin{equation}
 \mathcal{E}_{\mathcal{N},m}^{\pm}\simeq E^{e}_{\mathcal{N}}+E_{q}+m_{n}-m_{p}+(k_{z}^{2}+q^{2}+2\mathcal{N}eB)/2m_{d}+q_{z}k_{z}/m_{d}+m\omega\pm q_{y}\sqrt{2\mathcal{N}eB}/m_{d}.\label{SM13}
 \end{equation} 
As a function of $\mathcal{E}_{\mathcal{N},m}^{\pm}$, Eq.(\ref{SM11}) can be written as
\begin{align}
&\left[\frac{\langle\phi^{0}_{int}|e^{i(2/3)[\mathbf{r}_{\lambda}\cdot(\mathbf{k}+\mathbf{q})+r_{\lambda}^{y}(\sqrt{2\mathcal{N}eB}-k_{y})]}|\phi_{int}^{m}\rangle}{\mathcal{E}^{+}_{\mathcal{N},m}}+\frac{\langle\phi^{0}_{int}|e^{i(2/3)[\mathbf{r}_{\lambda}\cdot(\mathbf{k}+\mathbf{q})+r_{\lambda}^{y}(-\sqrt{2\mathcal{N}eB}-k_{y})]}|\phi_{int}^{m}\rangle}{\mathcal{E}^{-}_{\mathcal{N},m}}\right]\nonumber\\
&\times e^{-\xi^{2}_{e^{+}}/2}H_{\mathcal{N}}(\xi_{e^{+}})\frac{e^{i(p_{y}+q_{y})(y-y')}(\mathbf{k}+\mathbf{q})\cdot\hat{\mathbf{z}}}{4}e^{-\xi^{'2}_{e^{+}}/2}H_{\mathcal{N}}(\xi'_{e^{+}})\label{SM14}\\
&\times\left[\frac{\langle\phi^{m}_{int}|e^{-i(2/3)[\mathbf{r}_{\lambda}\cdot(\mathbf{k}+\mathbf{q})+r_{\lambda}^{y}(\sqrt{2\mathcal{N}eB}-k_{y})]}|\phi_{int}^{0}\rangle}{\mathcal{E}^{+}_{\mathcal{N},m}}+\frac{\langle\phi^{m}_{int}|e^{-i(2/3)[\mathbf{r}_{\lambda}\cdot(\mathbf{k}+\mathbf{q})+r_{\lambda}^{y}(-\sqrt{2\mathcal{N}eB}-k_{y})]}|\phi_{int}^{0}\rangle}{\mathcal{E}^{-}_{\mathcal{N},m}}\right].\nonumber
\end{align}
In this equation we identify the form factors between the wavefunctions of the internal nucleon states $0$ and $m$,
\begin{equation}
\langle\phi^{0}_{int}|e^{\pm i(2/3)[\mathbf{r}_{\lambda}\cdot(\mathbf{k}+\mathbf{q})+r_{\lambda}^{y}(\pm\sqrt{2\mathcal{N}eB}-k_{y})]}|\phi_{int}^{m}\rangle,\label{formfactors}\\
\end{equation}
which have been denoted by $\langle\tilde{\phi}^{0}_{int}|\tilde{\phi}_{int}^{m}\rangle$ in Eqs.(\ref{f1a}) and (\ref{f2a}) of Secs.\ref{Adiabatic} and \ref{spinrelax}.   Finally, approximating $\mathcal{E}_{\mathcal{N},m}^{+}\simeq\mathcal{E}_{\mathcal{N},m}^{-}\simeq\mathcal{E}_{\mathcal{N},m}=E^{e}_{\mathcal{N}}+E_{q}+m_{n}-m_{p}+(k_{z}^{2}+q^{2}+2\mathcal{N}eB)/2m_{d}+q_{z}k_{z}/m_{d}+m\omega$,  Eq.(\ref{SM11}) reads
\begin{align}
e^{i(p_{y}+q_{y})(y-y')}e^{-(\xi^{2}_{e^{+}}+\xi^{'2}_{e^{+}})/2}H_{\mathcal{N}}(\xi_{e^{+}})H_{\mathcal{N}}(\xi'_{e^{+}})\Big|\langle\tilde{\phi}^{0}_{int}|\tilde{\phi}_{int}^{m}\rangle\Big|^{2}\frac{k_{z}+q_{z}}{\mathcal{E}^{2}_{\mathcal{N},m}},\label{SM15}
\end{align} 
as it appears in Eqs.(\ref{f1a}) and (\ref{f2a}).

\subsection{Form factors $\langle\tilde{\phi}^{0}_{int}|\tilde{\phi}_{int}^{m}\rangle$}\label{AppD2}

Regarding the form factors of the internal states and the contributions of these states to the energy denominator $\mathcal{E}_{\mathcal{N},m}$,  we first note that the $\mathbf{r}_{\lambda}$-dependent and $\mathbf{r}_{\rho}$-dependent parts of the Hamiltonian $H_{int}^{N}$ in Eq.(\ref{Hint}) are isotropic harmonic oscillators of frequency $\omega$ and effective masses $2m_{d}/3$ and $m_{d}/2$, respectively. As for the sum over internal nucleon states in Eqs.(\ref{f1a}) and (\ref{f2a}), the states $|\phi^{m}_{int}\rangle$ are composed of the tensor product of the ground state of the oscillator in $\mathbf{r}_{\rho}$, and all the states of the oscillator in $\mathbf{r}_{\lambda}$. As for the wavefunction of the non-degenerate ground state, it reads
\begin{equation}
\varphi_{int}^{0}(\mathbf{r}_{\lambda},\mathbf{r}_{\rho})=\langle \mathbf{r}_{\lambda},\mathbf{r}_{\rho}|\phi_{int}^{0}\rangle=\frac{\beta^{3}}{3^{3/4}\pi^{3/2}}e^{-\beta^{2}(r_{\lambda}^{2}/3+r_{\rho}^{2}/4)},\quad\beta=\sqrt{\omega m_{d}},\label{phinot}
\end{equation}
while the excited eigenstates with energy $m\omega$, $m\in\mathbb{N}$, are degenerate with degree of degeneracy $g_{m}=(m+1)(m+2)/2$ and wavefunctions
\begin{equation}
\varphi_{int}^{m_{x},m_{y},m_{z}}(\mathbf{r}_{\lambda},\mathbf{r}_{\rho})=\frac{\varphi_{int}^{0}(\mathbf{r}_{\lambda},\mathbf{r}_{\rho})}{\sqrt{2^{m}m_{x}!m_{y}!m_{z}!}}H_{m_{x}}(\sqrt{2/3}\beta r_{\lambda}^{x})H_{m_{y}}(\sqrt{2/3}\beta r_{\lambda}^{y})H_{m_{z}}(\sqrt{2/3}\beta r_{\lambda}^{z}),\quad m\equiv m_{x}+m_{y}+m_{z}.
\end{equation}
Inserting the above expression in Eq.(\ref{formfactors}) for the form factors, and squaring, we end up with
\begin{equation}
|\langle\tilde{\phi}^{0}_{int}|\tilde{\phi}_{int}^{m}\rangle|^{2}=\frac{(3/4)^{m}}{m_{x}!m_{y}!m_{z}!}(\kappa_{x}/\beta)^{2m_{x}}(\kappa_{y}/\beta)^{2m_{y}}(\kappa_{z}/\beta)^{2m_{z}}e^{-3(\kappa_{x}^{2}+\kappa_{y\pm}^{2}+\kappa_{z}^{2})+/4\beta^{2}},\quad m= m_{x}+m_{y}+m_{z},\label{formy}
\end{equation}
where $\kappa_{x}=2(k_{x}+q_{x})/3$, $\kappa_{z}=2(k_{z}+q_{z})/3$, $\kappa_{y\pm}=2(q_{y}\pm\sqrt{2\mathcal{N}eB})/3$. On the other hand, the energy of the excited sates $m\omega$ becomes rapidly relativistic, since $\omega\sim m_{d}$. Therefore, in an effective manner, the combination $|\langle\tilde{\phi}^{0}_{int}|\tilde{\phi}_{int}^{m}\rangle|^{2}/\mathcal{E}_{\mathcal{N},m}^{2}$ in the integrand of Eqs.(\ref{f1a}) and (\ref{f2a}) scales rapidly as 
\begin{equation}
|\langle\tilde{\phi}^{0}_{int}|\tilde{\phi}_{int}^{m}\rangle|^{2}/\mathcal{E}_{\mathcal{N},m}^{2}\approx\frac{(3/4)^{m}}{m_{x}!m_{y}!m_{z}!m^{2}\omega^{2}}(\kappa_{x}/\beta)^{2m_{x}}(\kappa_{y}/\beta)^{2m_{y}}(\kappa_{z}/\beta)^{2m_{z}}e^{-3(\kappa_{x}^{2}+\kappa_{y\pm}^{2}+\kappa_{z}^{2})+/4\beta^{2}},\quad m= m_{x}+m_{y}+m_{z}.\label{formy1}
\end{equation}
In the first place, we note from Eqs.(\ref{formy}) and (\ref{formy1}) that the combination $|\langle\tilde{\phi}^{0}_{int}|\tilde{\phi}_{int}^{m}\rangle|^{2}/\mathcal{E}_{\mathcal{N},m}^{2}$ suppresses exponentially the integrand of the momentum integrals in Eqs.(\ref{f1a}) and (\ref{f2a}) for all $m$. It provides this way an effective cutoff at $q,k,\sqrt{2\mathcal{N}eB}\approx\beta\sim m_{d}$, which makes the integrals convergent. Second, the form factor of the excited state $m$ provides a factor which scales as $\sim(k^{2}+q^{2}+2\mathcal{N}eB)^{m}/m_{d}^{2m}$, which is a relativistic correction $\forall$ $m\geq1$ that lies beyond the non-relativistic approximation assumed for the internal hadron dynamics. Therefore, the contribution of all the excited states must be discarded by consistency. Nonetheless, one must bear in mind that their contribution would be just one order higher than the leading order nonrelativistic terms included in Eqs.(\ref{f1b}) and (\ref{f2b}). With this proviso, we are just left with the contribution of the internal nucleon ground state whose form factor reads
\begin{align}
\langle\phi^{0}_{int}|e^{\pm i(2/3)[\mathbf{r}_{\lambda}\cdot(\mathbf{k}+\mathbf{q})+r_{\lambda}^{y}(\pm\sqrt{2\mathcal{N}eB}-k_{y})]}|\phi_{int}^{0}\rangle&=\int\textrm{d}^{3}r_{\rho}\int\textrm{d}^{3}r_{\lambda}e^{\pm i(2/3)[\mathbf{r}_{\lambda}\cdot(\mathbf{k}+\mathbf{q})+r_{\lambda}^{y}(\pm\sqrt{2\mathcal{N}eB}-k_{y})]}\frac{\beta^{6}}{(\sqrt{3}\pi)^{3}}e^{-\beta^{2}(2r_{\lambda}^{2}/3+r_{\rho}^{2}/2)}\nonumber\\
&=e^{-[(k_{x}+q_{x})^{2}+(k_{z}+q_{z})^{2}+q_{y}^{2}+2\mathcal{N}eB]/6\beta^{2}}e^{\mp q_{y}\sqrt{2\mathcal{N}eB}/3\beta^{2}}.\label{SM16}
\end{align}
Substituting this formula into Eq.(\ref{SM14}) we end up with
\begin{align}
&(1/2)e^{i(p_{y}+q_{y})(y-y')}e^{-(\xi^{2}_{e^{+}}+(\xi')^{2}_{e^{+}})/2}H_{\mathcal{N}}(\xi_{e^{+}})H_{\mathcal{N}}(\xi'_{e^{+}})e^{-[(k_{x}+q_{x})^{2}+(k_{z}+q_{z})^{2}+q_{y}^{2}+2\mathcal{N}eB]/3\beta^{2}}\nonumber\\
&\times(\mathbf{k}+\mathbf{q})\left[\frac{e^{-2q_{y}\sqrt{2\mathcal{N}eB}/3\beta^{2}}}{(\mathcal{E}^{+}_{\mathcal{N}})^{2}}+\frac{1}{\mathcal{E}^{+}_{\mathcal{N}}\mathcal{E}^{-}_{\mathcal{N}}}\right],\nonumber
\end{align}
where it is assumed that $\sqrt{2\mathcal{N}eB}$ takes both positive and negative values. %As advanced in the Letter, the Fourier transform of the square of the wavefunction is a Gaussian function which, in an effective manner, cuts off the momentum integrals at $k,q,\sqrt{2\mathcal{N}eB}\approx\alpha\approx m_{d}$.

\section{Magnetic nature of the proton kinetic energy}\label{AppE}

In this Appendix we demonstrate the magnetic nature of the kinetic energy gained by the proton along the process described in Sec.\ref{Adiabatic}.\\

Let us consider first the variation of the total energy of the proton with respect to variations of the magnetic field strength, $\delta B$, at certain value of the field $\mathbf{B}$,
%\begin{equation}
%\langle\delta H/\delta B\rangle_{B}=\left\langle\delta\left[\sum_{\alpha}H_{D}^{\alpha}+W+\delta T\right]/\delta B\right\rangle_{B}=\langle [H_{D}^{p}+H_{D}^{n}+H_{D}^{e}]/B\rangle_{B},\nonumber\qquad\textrm{(FA1)}
%\end{equation}
\begin{equation}
\langle\delta H/\delta B\rangle_{\mathbf{B}}=\langle\delta[H_{0}+W]/\delta B\rangle_{\mathbf{B}}=\langle\delta[H_{mag}^{p}+H_{mag}^{n}+H_{mag}^{e}]/\delta B\rangle_{\mathbf{B}},\nonumber\qquad
\end{equation}
where the total Hamiltonian is evaluated in the state of the system at a certain value of the field $\mathbf{B}$, which is denoted by $\rangle_{\mathbf{B}}$. %In the last equality I have made use of the fact that only the magnetic terms of the Dirac Hamiltonians of protons, electrons and neutrons depend on $B$, and that dependence is linear. 
Taking into account that $B$ is an adiabatic parameter, $\delta B$ can be factored out of the expectation values of the above equation,
\begin{equation}
\langle\delta H/\delta B\rangle_{\mathbf{B}}=\langle\delta[H_{mag}^{p}+H_{mag}^{n}+H_{mag}^{e}]/\delta B\rangle_{\mathbf{B}}=\delta\left\langle\sum_{\alpha}H_{D}^{\alpha}+W+\delta T\right\rangle_{\mathbf{B}}/\delta B.\label{SM17}
\end{equation}
Next, we will show that the variation of the kinetic energy of the proton is indeed part of the variation of the magnetic energy, % $\left\langle\sum_{\alpha}H_{D}^{\alpha}+W+\delta T\right\rangle_{B}/B$, 
 since it is one of the terms within $\delta\langle W\rangle_{\mathbf{B}}/\delta B$. That is, we will prove that 
\begin{align}
&\frac{\delta\langle\mathbf{P}_{N}\rangle_{\mathbf{B}}^{2}}{2m_{p}\delta B}\textrm{ is part of }\delta\langle W\rangle_{\mathbf{B}}/\delta B,\textrm{ which is part of }\langle\delta H/\delta B\rangle_{\mathbf{B}}=\langle \delta[H_{mag}^{p}+H_{mag}^{n}+H_{mag}^{e}]/\delta B\rangle_{\mathbf{B}},\nonumber\\
&\textrm{from which it holds that }\frac{\delta\langle\mathbf{P}_{N}\rangle_{\mathbf{B}}^{2}}{2m_{p}\delta B}\textrm{ is part of }\langle \delta[H_{mag}^{p}+H_{mag}^{n}+H_{mag}^{e}]/\delta B\rangle_{\mathbf{B}}.\label{SM18}%\left\langle\sum_{\alpha}H_{D}^{\alpha}+W+\delta T\right\rangle_{B}/B
\end{align}
Here $\langle\mathbf{P}_{N}\rangle_{\mathbf{B}}$ is the momentum of the proton at magnetic field $\mathbf{B}$, computed according to Eq.(\ref{equationmother}), which is linear in $\mathbf{B}$. 

Let us start with the expression for $\langle W\rangle_{\mathbf{B}'}$ at a quasi-stationary field $\mathbf{B}(t')\equiv\mathbf{B}'=\mathbf{B}+\delta\mathbf{B}$ whose diagrammatic representation is that of Fig.\ref{SMfig}, at $\mathcal{O}(W^{2})$,
\begin{align}
\langle W\rangle_{\mathbf{B}'}&=_{\mathbf{B}'}\langle\tilde{\Phi}|\tilde{\mathbb{U}}^{\dagger}_{\mathbf{B}'}(t')\:W\:\tilde{\mathbb{U}}_{\mathbf{B}'}(t')|\tilde{\Phi}\rangle_{\mathbf{B}'}\nonumber\\
&=-\frac{G_{F}^{2}}{2}\textrm{Re}\int\textrm{d}^{3}R\textrm{d}^{3}R'\:_{\mathbf{B}'}\langle\tilde{\Phi}|J_{h}^{\mu}(\mathbf{R},\tilde{\mathbf{r}})J^{l}_{\mu}(\tilde{\mathbf{r}})\frac{1}{H_{0}-E_{0}^{p}}
J^{\rho\dagger}_{\mu}(\tilde{\mathbf{r}}')|J_{h}^{\rho\dagger}(\mathbf{R}',\tilde{\mathbf{r}}')|\tilde{\Phi}\rangle_{\mathbf{B}'},\quad\mu,\rho=0,..,3,\label{SM19}
\end{align}
%\begin{multicols}{2}
\begin{figure}[h!]
\centering
\includegraphics[height=6.3cm,width=6.cm,clip]{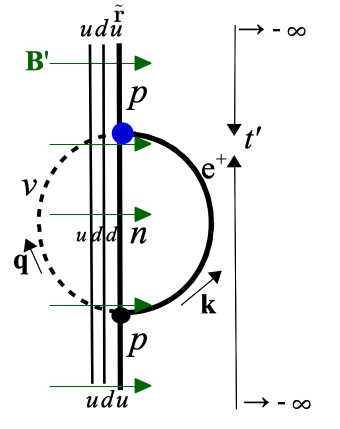}
\caption{Diagrammatic representation of $\langle W\rangle_{\mathbf{B}'}$ for a quasi-stationary magnetic field $\mathbf{B}'$, according to Eq.(\ref{SM19}). The calculation is performed in the adiabatic limit, where the initial time is taken to $-\infty$ according to Eq.(\ref{eta}), and the operator $W$, represented with a blue circle, applies at the observation time $t'$.}\label{SMfig}
\end{figure}
where $|\tilde{\Phi}\rangle_{\mathbf{B}'}=|\Omega_{l}\rangle_{\mathbf{B}'}\otimes|\phi_{int}^{N}\rangle\otimes\big[\big|0,1;\langle\mathbf{P}_{N}\rangle_{\mathbf{B}}\big\rangle+\big|0,-1;\langle\mathbf{P}_{N}\rangle_{\mathbf{B}}\big\rangle\big]^{p}_{\mathbf{B}'}/2L\sqrt{E^{p}_{0}}$ is the normalized state of the system in a magnetic field $\mathbf{B}'$, with momentum $\langle\mathbf{P}_{N}\rangle_{\mathbf{B}}$ along $\mathbf{B}$, and $E_{0}^{p}$ incorporates its kinetic energy $|\langle\mathbf{P}_{N}\rangle_{\mathbf{B}}|^{2}/2m_{p}$.  Denoting by $\mathbf{R}$ the centre of mass vector of the proton, we can write instead, $|\tilde{\Phi}\rangle_{\mathbf{B}'}=|\Omega_{l}\rangle_{\mathbf{B}'}\otimes|\phi_{int}^{N}\rangle\otimes e^{i\langle\mathbf{P}_{N}\rangle_{\mathbf{B}}\cdot\mathbf{R}}\big[|0,1;\mathbf{0}\rangle+|0,-1;\mathbf{0}\rangle\big]^{p}_{\mathbf{B}'}/2L\sqrt{E^{p}_{0}}$.  Note also that $\mathbf{B}'$ and $\mathbf{B}$ differ by $\delta\mathbf{B}$. This will allow us to identify $\delta\langle W\rangle_{\mathbf{B}}$ from the relationship $\langle W\rangle_{\mathbf{B}'}=\langle W\rangle_{\mathbf{B}}+\delta\langle W\rangle_{\mathbf{B}}$.
In Appendix \ref{AppD} we computed the energy factor $\mathcal{E}_{\mathcal{N},m}$ starting with the expression in Eq.(\ref{SM9}).  In the present case, that equation must incorporate the factor $e^{-i\langle\mathbf{P}_{N}\rangle_{\mathbf{B}}\cdot\mathbf{R}}$ of $\:_{\mathbf{B}'}\langle\tilde{\Phi}|$ on the left, and $e^{i\langle\mathbf{P}_{N}\rangle_{\mathbf{B}}\cdot\mathbf{R}'}$ on the right of the integrand,
\begin{align}
&\int\frac{\textrm{d}x\textrm{d}z\textrm{d}p_{x}\textrm{d}p_{z}}{(2\pi)^{2}}\langle\phi^{0}_{int}|e^{i\mathbf{R}\cdot(\mathbf{p}+\mathbf{k}+\mathbf{q}-\langle\mathbf{P}_{N}\rangle_{\mathbf{B}})-ik_{y}R_{y}}e^{i(2/3)[\mathbf{r}_{\lambda}\cdot(\mathbf{k}+\mathbf{q})-k_{y}r_{\lambda}^{y}]}e^{-\xi^{2}_{e^{+}}/2}H_{\mathcal{N}}(\xi_{e^{+}})\cos{(2r^{y}_{\lambda}\sqrt{2\mathcal{N}eB}/3)}\nonumber\\
&\times|\phi_{int}^{m}\rangle\frac{1}{E^{e}_{\mathcal{N}}+E_{q}+E^{n}_{\mathcal{S}}(\mathbf{p})-E_{0}^{p}+H_{int}^{N}(\mathbf{p}_{\rho};\mathbf{p}_{\lambda})}\langle\phi_{int}^{m}|\nonumber\\
&\times e^{-\xi^{'2}_{e^{+}}/2}H_{\mathcal{N}}(\xi'_{e^{+}})\cos{(2r^{y}_{\lambda}\sqrt{2\mathcal{N}eB}/3)} e^{-i\mathbf{R}'\cdot(\mathbf{p}+\mathbf{k}+\mathbf{q}+\langle\mathbf{P}_{N}\rangle_{\mathbf{B}})+ik_{y}R'_{y}}e^{-i(2/3)[\mathbf{r}_{\lambda}\cdot(\mathbf{k}+\mathbf{q})-k_{y}r_{\lambda}^{y}]}|\phi_{int}^{0}\rangle.\label{SM20}
\end{align}
In contrast to Eq.(\ref{SM9}), the integration in $z$ of Eq.(\ref{SM20}) yields now $2\pi\delta(p_{z}+q_{z}+k_{z}-\langle P_{N}\rangle_{\mathbf{B}})$, from which the energy factor of Eq.(\ref{SM13}) picks up an additional term, 
\begin{equation}
\mathcal{E}^{\pm}_{\mathcal{N},m}\rightarrow \mathcal{E}^{\pm}_{\mathcal{N},m}-(\mathbf{k}+\mathbf{q})\cdot\langle\mathbf{P}_{N}\rangle_{\mathbf{B}}/m_{p}.\quad\nonumber
\end{equation}
The additional term is the energy associated to the Doppler shift of the virtual positrons and neutrinos which interact with the proton in motion. Following steps analogous to those in Appendix \ref{AppD}, Eq.(\ref{SM20}) can be written as
\begin{align}
\frac{e^{i(p_{y}+q_{y})(y-y')}e^{-(\xi^{2}_{e^{+}}+(\xi')^{2}_{e^{+}})/2}H_{\mathcal{N}}(\xi_{e^{+}})H_{\mathcal{N}}(\xi'_{e^{+}})\Big|\langle\tilde{\phi}^{0}_{int}|\tilde{\phi}_{int}^{m}\rangle\Big|^{2}}{\mathcal{E}_{\mathcal{N},m}-(\mathbf{k}+\mathbf{q})\cdot\langle\mathbf{P}_{N}\rangle_{\mathbf{B}}/m_{p}}.\label{SM21}
\end{align} 
Finally, expanding the above equation up to leading order in $(\mathbf{k}+\mathbf{q})\cdot\langle\mathbf{P}_{N}\rangle_{\mathbf{B}}/m_{p}\mathcal{E}_{\mathcal{N},m}$, Eq.(\ref{SM21}) is approximately equal to
\begin{align}
e^{i(p_{y}+q_{y})(y-y')}e^{-(\xi^{2}_{e^{+}}+(\xi')^{2}_{e^{+}})/2}H_{\mathcal{N}}(\xi_{e^{+}})H_{\mathcal{N}}(\xi'_{e^{+}})\Big|\langle\tilde{\phi}^{0}_{int}|\tilde{\phi}_{int}^{m}\rangle\Big|^{2}\left[\frac{1}{\mathcal{E}_{\mathcal{N},m}}+\frac{(\mathbf{k}+\mathbf{q})\cdot\langle\mathbf{P}_{N}\rangle_{\mathbf{B}}/m_{p}}{\mathcal{E}^{2}_{\mathcal{N},m}}\right].\label{SM22}
\end{align}
The first term does not depend on the proton momentum, whereas the second term is exactly Eq.(\ref{SM15}) found in Appendix \ref{AppD} for the Casimir momentum times $\langle\mathbf{P}_{N}\rangle_{\mathbf{B}}/m_{p}$. Putting all these results together, Eq.(\ref{SM19}) yields, up to terms of order $(\mathbf{k}+\mathbf{q})\cdot\langle\mathbf{P}_{N}\rangle_{\mathbf{B}}/m_{p}\mathcal{E}_{\mathcal{N},m}$ in the integrand,
\begin{align}
\langle&W\rangle_{\mathbf{B}'}=-\frac{G_{F}^{2}}{2}\textrm{Re}\int\textrm{d}^{3}R\textrm{d}^{3}R'\:_{\mathbf{B}'}\langle\Phi|J_{h}^{\mu}(\mathbf{R},\tilde{\mathbf{r}})J^{l}_{\mu}(\tilde{\mathbf{r}})\frac{1}{H_{0}-E_{0}^{p}}
J^{\rho\dagger}_{\mu}(\tilde{\mathbf{r}}')J_{h}^{\rho\dagger}(\mathbf{R}',\tilde{\mathbf{r}}')|\Phi\rangle_{\mathbf{B}'}\nonumber\\
&-\frac{G_{F}^{2}}{2}\textrm{Re}\int\textrm{d}^{3}R\textrm{d}^{3}R'\:_{\mathbf{B}'}\langle\Phi|J_{h}^{\mu}(\mathbf{R},\tilde{\mathbf{r}})J^{l}_{\mu}(\tilde{\mathbf{r}})\frac{(\mathbf{K}_{e}+\mathbf{Q}_{\nu})\cdot\langle\mathbf{P}_{N}\rangle_{B}/m_{p}}{[H_{0}-E_{0}^{p}]^{2}}
J^{\rho\dagger}_{\mu}(\tilde{\mathbf{r}}')J_{h}^{\rho\dagger}(\mathbf{R}',\tilde{\mathbf{r}}')|\Phi\rangle_{\mathbf{B}'},\label{SM23}
\end{align}
where $|\Phi\rangle_{\mathbf{B}'}=|\Omega_{l}\rangle_{\mathbf{B}'}\otimes|\phi_{int}^{N}\rangle\otimes\big[\big|0,1;\mathbf{0}\big\rangle+\big|0,-1;\mathbf{0}\big\rangle\big]^{p}_{\mathbf{B}'}/2L\sqrt{E^{p}_{0}}$ is here the state of the system in a magnetic field $\mathbf{B}'$, where the proton presents zero momentum along $\mathbf{B}$. The first term of Eq.(\ref{SM23}) is the leading order EW self-energy of a proton in the presence of a constant and uniform magnetic field $\mathbf{B}'$, while the second term is the leading order Doppler-shift correction. By direct comparison with Eq.(\ref{equationmother}), the second term is readily identifiable with $-\langle\mathbf{K}_{e^{+}}+\mathbf{Q}_{\nu}\rangle_{\mathbf{B}'}\cdot\langle\mathbf{P}_{N}\rangle_{\mathbf{B}}/m_{p}$. As argued in Sec.\ref{Adiamomentum}, conservation of total momentum along $\mathbf{B}'$ implies $-\langle\mathbf{K}_{e^{+}}+\mathbf{Q}_{\nu}\rangle_{\mathbf{B}'}=\langle\mathbf{P}_{N}\rangle_{\mathbf{B}'}$. Lastly, since $\mathbf{B}'=\mathbf{B}+\delta\mathbf{B}$, we can write the nucleon momentum at $\mathbf{B}'$ as $\langle\mathbf{P}_{N}\rangle_{\mathbf{B}'}=\langle\mathbf{P}_{N}\rangle_{\mathbf{B}}+\delta\langle\mathbf{P}_{N}\rangle_{\mathbf{B}}$, from which we obtain that 
\begin{equation}
\frac{\langle\mathbf{P}_{N}\rangle_{\mathbf{B}'}\cdot\langle\mathbf{P}_{N}\rangle_{\mathbf{B}}}{m_{p}}=\frac{\langle\mathbf{P}_{N}\rangle_{\mathbf{B}}^{2}}{m_{p}}+\frac{\delta\langle\mathbf{P}_{N}\rangle_{\mathbf{B}}\cdot\langle\mathbf{P}_{N}\rangle_{\mathbf{B}}}{m_{p}}\textrm{ is part of }\langle W\rangle_{\mathbf{B}'}=\langle W\rangle_{\mathbf{B}}+\delta\langle W\rangle_{\mathbf{B}}.\label{SM24}
\end{equation}
Identifying $\delta\langle\mathbf{P}_{N}\rangle_{\mathbf{B}}\cdot\langle\mathbf{P}_{N}\rangle_{\mathbf{B}}/m_{p}$ as part of $\delta\langle W\rangle_{\mathbf{B}}$ in Eq.(\ref{SM24}), %and identifying
%\begin{equation}
%-\langle\dot{\mathbf{K}}_{e^{+}}+\dot{\mathbf{Q}}_{\nu}\rangle_{\mathbf{B}}=\langle\dot{\mathbf{P}}_{N}\rangle_{\mathbf{B}}=\frac{\delta\langle\mathbf{P}_{N}\rangle_{\mathbf{B}}}{\delta\mathbf{B}}\frac{\partial\mathbf{B}}{\partial t}\equiv\langle\mathbf{F}\rangle_{\mathbf{B}}\label{SM25}
%\end{equation}
%with the force acting on the proton, 
we conclude the proof of the relationship (\ref{SM18}) and confirm the statements in Sec.\ref{AdiaEnergy}.

\section{Some integrals and sums of infinite series}\label{AppF}

The integrals in Eqs.(\ref{f1a}) and (\ref{f2a}) involve a number of technical issues. Here we summarize the most relevant ones.

As for the integrals involving $I_{\mathcal{N}}$ functions, the following identities hold.
\begin{align}
&\int_{-\infty}^{\infty}\textrm{d}y\textrm{d}y'\int_{-\infty}^{\infty}\frac{\textrm{d}p_{y}\textrm{d}k_{x}}{(2\pi)^{2}}e^{ip_{y}(y-y')}I_{\mathcal{N}}(y)I_{\mathcal{N}}(y')I_{\mathcal{N}'}(\xi_{e^{+}})I_{\mathcal{N}'}(\xi'_{e^{+}})=eB/2\pi,\nonumber\\
&\int_{-\infty}^{\infty}\textrm{d}y\textrm{d}y'\int_{-\infty}^{\infty}\frac{\textrm{d}p_{y}\textrm{d}k_{x}}{2(2\pi)^{2}}e^{ip_{y}(y-y')}k_{x}\sqrt{2\mathcal{N}eB}I_{0}(y)I_{0}(y')[I_{\mathcal{N}}(\xi_{e^{+}})I_{\mathcal{N}-1}(\xi'_{e^{+}})+I_{\mathcal{N}}(\xi'_{e^{+}})I_{\mathcal{N}-1}(\xi_{e^{+}})]=\mathcal{N}e^{2}B^{2}/\pi.\nonumber
\end{align}

Regarding the sums over infinite series, considering that $eB\ll m_{d}^{2}$ and that the range of $k_{z}$ for which the integrands yield relevant contributions to the momentum integrals is such that $|k_{z}|\lessapprox m_{d}$, we perform the passage to the continuum by means of the following substitutions,
\begin{equation}
E^{2}_{\mathcal{N}B}\equiv2\mathcal{N}eB,\qquad\sum_{\mathcal{N}=0}^{\infty}f(E_{\mathcal{N}B})\rightarrow\int_{0}^{\infty}\frac{E_{\mathcal{N}B}\textrm{d}E_{\mathcal{N}B}}{eB}f(E_{\mathcal{N}B}),\quad\textrm{for any function }f\textrm{ of }E_{\mathcal{N}B}.\nonumber
\end{equation}  

\end{widetext}


\begin{thebibliography}{105}
\bibitem{Miltonbook} K.A. Milton, \emph{The Casimir Effect: Physical Manifestations of Zero-point Energy}, World Sci., Singapore (2001).
\bibitem{Milonnibook}  P.W. Milonni,  \textit{The Quantum Vacuum}, Academic Press, San Diego (1994).
\bibitem{Weinberg} S.  Weinberg,  \textit{The cosmological constant problem}, Rev.  Mod. Phys. \textbf{61}, 1 (1989).
\bibitem{Jaffe} R.L. Jaffe, \textit{Casimir effect and the quantum vacuum}, Phys. Rev. D \textbf{72}, 021301(R) (2005).
\bibitem{Casimir} H.B.G. Casimir, Proc. K. Ned. Akad. Wet. \textbf{51}, 793 (1948); E. M. Lifschitz, \textit{The theory of molecular attractive forces between solids}, Sov. Phys. \textbf{2}, 73 (1956); S.K. Lamoreaux,  \textit{Demonstration of the Casimir Force in the 0.6 to 6$\mu$m Range}, Phys. Rev. Lett. \textbf{78}, 5 (1997), Erratum: Phys. Rev. Lett. \textbf{81}, 5475 (1998). 
\bibitem{vdW} F. London, \textit{Zur Theorie und Systematik der Molekularkr\"{a}fte}, Z. Phys. \textbf{63}, 245 (1930); H.B.G. Casimir  and D. Polder, \textit{The Influence of Retardation on the London-van der Waals Forces}, \emph{Phys. Rev.} \textbf{73}, 360 (1948).
 \bibitem{Bethe} W.E. Lamb and R.C. Retherford, \textit{Fine Structure of the Hydrogen Atom by a Microwave Method}, Phys. Rev. \textbf{72}, 241(1947); H. A. Bethe, \textit{The Electromagnetic Shift of Energy Levels}, Phys. Rev. \textbf{72}, 339 (1947).
 \bibitem{cavitychromo} A. Chodos, R.L. Jaffe, K. Johnson, C.B Thorn and V.F. Weisskopf, \textit{New Extended Model of Hadrons}, Phys. Rev. D \textbf{9}, 3471 (1974);  P. Hasenfratz and J. Kuti, \textit{The quark bag model}, Phys. Rep. \textbf{40}, 75 (1978). 
 \bibitem{Cherenkov} M.N. Chernodub, V.A. Goy, A.V. Molochkov, and H.H. Nguyen, \textit{Casimir Effect in Yang-Mills Theory in $D=2+1$}, Phys. Rev. Lett. \textbf{121}, 191601 (2018).
\bibitem{Feigel} A. Feigel, \textit{Quantum Vacuum Contribution to the Momentum of Dielectric Media}, Phys. Rev. Lett. \textbf{92}, 020404 (2004).
\bibitem{Croze}  B.A. van Tiggelen, G.L.J.A. Rikken and V. Krsti\'c, \textit{Momentum Transfer from Quantum Vacuum to Magnetoelectric Matter},  Phys. Rev. Lett. \textbf{96}, 130402 (2006); B.A. van Tiggelen, \textit{Zero-point momentum in complex media}, Eur. Phys. J. D \textbf{47}, 261 (2008);  B.A. van Tiggelen, S. Kawka and G.L.J.A. Rikken, \textit{QED Corrections to the Electromagnetic Abraham Force. Casimir Momentum of the Hydrogen Atom?}, Eur. Phys. J. D \textbf{66}, 272  (2012); O.A. Croze, \textit{Alternative derivation of the Feigel effect and call for its experimental verification}, Proc. R. Soc. A \textbf{468}, 429 (2012). 
\bibitem{PRLDonaire} M. Donaire, B.A. van Tiggelen and G.L.J.A. Rikken, \textit{Casimir Momentum of a Chiral Molecule in a Magnetic Field}, Phys. Rev. Lett. {\bf 111}, 143602 (2013).
\bibitem{JPCMDonaire} M. Donaire, B.A. van Tiggelen and G.L.J.A. Rikken, \textit{Transfer of linear momentum from the quantum vacuum to a magnetochiral molecule}, \emph{J. Phys.: Condens. Matter}  {\bf 27}, 214002 (2015).
%\bibitem{remark} We consider that the nucleon interacts with the magnetic field as a whole. Despite the  coupling between the spins of the quarks \cite{deRujula,Coince,Isgur}, their total spin contribution to the magnetic moment of the nucleons is equal to the magnetic moment of a single quark \cite{Karliner}.
%\bibitem{SM} See Supplemental Material at http://URL for a compilation of the spinor eigenstates of $H_{D}^{\alpha}+H_{mag}^{\alpha}$ for each EW particle $\alpha$. It also comprises a derivation of the energy factor $\mathcal{E}_{\mathcal{N}}$ and the square of the wavefunction $\langle\tilde{\phi}_{int}^{N}|\tilde{\phi}_{int}^{N}\rangle$ in Eqs.(\ref{f1a}) and (\ref{f2a}); the expressions for the hadronic and leptonic current fluctuations which enter Eq.(\ref{equationmother}); as well as the proof of the magnetic character of the kinetic energy.
\bibitem{Coince}  L. Conci and M. Traini,  \textit{Quark Momentum Distribution in Nucleons}, Few-Body Systems , \textbf{8},  123 (1990).
\bibitem{IsgurKarl1978}  N. Isgur and G. Karl,  \textit{P-wave baryons in the quark model}, Phys. Rev. D \textbf{18}, 4187 (1978).
\bibitem{IsgurKarl1979}  N. Isgur and G. Karl,  \textit{Positive-parity excited baryons in a quark model with hyperfine interactions}, Phys. Rev. D \textbf{19}, 2653 (1979).
\bibitem{deRujula}  A. de R\'ujula, H. Georgi and S.L. Glashow,  \textit{Hadron masses in a gauge theory}, Phys. Rev. D \textbf{12}, 147 (1975).
\bibitem{Isgur}  N. Isgur,  \textit{Mesonlike baryons and the spin-orbit puzzle}, Phys. Rev. D\textbf{62}, 014025 (2000).
\bibitem{Karliner} M. Karliner and H.J. Lipkin, \textit{New quark relations for hadron masses and magnetic moments: A challenge for explanation from QCD}, Phys. Lett. B \textbf{650}, 185 (2007).
\bibitem{Scadron} M.D. Scadron, R. Delbourgo, and G. Rupp, \textit{Constituent Quark Masses and the Electroweak Standard Model}, J. Phys. G: Nucl. Part. Phys. \textbf{32}, 736 (2006).
 \bibitem{Mao} G.J. Mao, A. Iwamoto and Z.X.Li, \textit{A Study of Neutron Star Structure in Strong Magnetic Fields that includes Anomalous Magnetic Moments}, Chin. J. Astron. Astrophys. \textbf{3}, 359 (2003); H. Wen, L.S. Kisslinger, W. Greiner and J.G. Mao, \textit{Neutron spin polarization in strong magnetic fields}, Int. J. Mod. Phys. E \textbf{14}, 1197 (2005).
\bibitem{Peskin} M.E. Peskin and D.V. Schroeder, \textit{An Introduction to Quantum Field Theory}, Westview Press, Chicago (1995); C. Itzykson and J.B. Zuber, \textit{Quantum Field Theory}, McGraw-Hill Inc. (1980); J. J. Sakurai, \textit{Advanced Quantum Mechanics}, Addison-Wesley, Boston (1994).
\bibitem{Bhattacharya} K. Bhattacharya and P.B. Pal,  Pramana - J. Phys.  \textbf{62}, 1041 (2004); K. Bhattacharya, \textit{Solution of the Dirac equation in presence of an uniform magnetic field}, arXiv:0705.4275 [hep-th] (2007).
\bibitem{Bander} M. Bander and H.R. Rubinstein, \textit{Proton Beta decay in large magnetic fields}, Phys. Lett. B \textbf{311}, 187 (1993).
 \bibitem{Broderick} A. Broderick, M. Prakash and J.M. Lattimer, \textit{The Equation of State of Neutron Star Matter in Strong Magnetic Fields}, ApJ \textbf{537}, 351 (2000).
\bibitem{FeynmanGellmann} R.P. Feynman and M. Gell-Mann,  \textit{Theory of the Fermi Interaction},  Phys. Rev. \textbf{109}, 193 (1958); 
E.C.G. Sudarshan and R. Marshak,  \textit{Chirality invariance and the Universal Fermi Interaction},  Phys. Rev. \textbf{109}, 1860 (1958).
 \bibitem{betadecay} J.D. Jackson, S.B. Treiman and H.W. Wyld. Jr., \textit{Possible Test of Time Reversal Invariance in Beta Decay}, Phys. Rev. \textbf{106}, 517 (1957); J.S. Nico, \textit{Neutron Beta Decay}, J. Phys. G: Nucl. Part. Phys. \textbf{36}, 104001 (2009); M.S. Sozzi, \textit{Discrete Symmetries and CP Violation. From Experiment to Theory}, Oxford University Press Inc., New York (2008).
%\bibitem{Morpurto} G. Morpurgo, \textit{Is a Non-relativistic Approximation possible for the Internal Dynamics of Elementary Particles?} 
%\bibitem{remark} The confining potentials, different for each nucleon, depend also on the individual spins of the quarks \cite{DeRujula,Coinci,Isgur}. 
 %However, by assuming that $\mathbf{B}_{0}$ couples only to the total spin of the nucleons, we neglect magnetic transitions between their internal states.
\bibitem{Wu} C.S. Wu, E. Ambler, R.W. Hayward, D.D. Hoppes, and R.P. Hudson, \textit{Experimental Test of Parity Conservation in Beta Decay}, Phys. Rev. 
 \textbf{105}, 1413 (1957).
  \bibitem{Pines} D. Pines and P. Nozieres, \emph{The Theory of Quantum Liquids}, Benjamin, New York (1966), p. 295-7.
% \bibitem{canonical} Although the component of the kinetic momentum of positrons along $\hat{\mathbf{x}}$ is $\sqrt{eB}\xi_{e^{+}}$, the inclusion of the canonical $k_{x}$ instead of the kinetic  $\sqrt{eB}\xi_{e^{+}}$ in $(\mathbf{k}+\mathbf{q})$ within the integrands of Eqs.(\ref{f1a},\ref{f2a}) is correct, since the expectation values of both momenta are null along $\hat{\mathbf{x}}$. 
%\bibitem{Jacksonbook} J.D. Jackson, \textit{Classical Electrodynamics}, 3rd ed., John Wiley \& Sons, Inc. (1988).
%\bibitem{HeroldRudder82}  H. Herold , H. Rudder and G. Wunner,  \textit{Cyclotron Emission in Strongly Magnetized Plasmas}, Astron. Astrophys. \textbf{115}, 90 (1981).
%\bibitem{Latal86}  H.G. Latal,  \textit{Cyclotron Radiation in Strong Magnetic Fields}, ApJ \textbf{309}, 372 (1986); M.G. Baring, P.L. Gonthier and A.K. Harding,  \textit{Spin-dependent Cyclotron Decay Rates in Strong Magnetic Fields}, ApJ \textbf{630}, 430 (2005).
\bibitem{Khalilov2005}  V.R. Khalilov,  \textit{Electroweak Nucleon Decays in a Superstrong Magnetic Field}, Theor. Math. Phys. \textbf{145}, 1462 (2005). 
\bibitem{Buhmann} S. Scheel  and  S.Y. Buhmann,  \emph{Acta Phys. Slov.} \textbf{58}  675 (2004).
\bibitem{MelroseAstronAstroph1981}  D.B. Melrose and V.V. Zheleznyakov,  \textit{Quantum Theory of Cyclotron Emission and the X-ray Line in Her X-1}, Astron. Astrophys. \textbf{95}, 86 (1981).
\bibitem{PothekinMonRoySoc2007}  A.Y. Pothekin and D. Lai,  \textit{Statistical Equilibrium and Ion Cyclotron Absorption/Emission in Strongly Magnetized Plasmas}, Mon. Not. R. Astron. Soc. \textbf{376}, 793 (2007).
\bibitem{Penning} K. Blaum and F. Herfurth, \textit{Trapped Charged Particles and Fundamental Interactions}, Springer-Verlag, Berlin-Heidelberg (2008); K. Blaum, Y.N. Novikov and G. Werth, \textit{Penning Traps as a versatile tool for Precise Experiments in Fundamental Physics}, Contemp. Phys. \textbf{51}, 149 (2010).
\bibitem{Griffiths} D. J Griffiths, \emph{Introduction to Electrodynamics}, Prentice-Hall, New Jersey (1999).
\bibitem{PRL106_253001(2011)} S. Ulmer, C.C. Rodegheri, K. Blaum, H. Kracke, A. Mooser, W. Quint, and J. Walz, \textit{Observation of Spin Flips with a Single Trapped Proton}, Phys. Rev. Lett. \textbf{106}, 253001 (2011); A. Mooser \emph{et al.}, \textit{Demonstration of the double Penning Trap technique with a single proton}, Phys. Lett. B \textbf{723}, 78 (2013); L.S. Brown, \textit{Geonium Lineshape}, Ann. Phys. (N.Y.) \textbf{159}, 62 (1985).
\bibitem{QuarkModels} D. Flamm and F. Schoberl, \textit{Introduction to the Quark Model of Elementary Particles}, Gordon \& Breach Science Pub. (1983); R.F. \'Alvarez-Estrada, F. Fern\'andez, J.L. S\'anchez-G\'omez and V. Vento, \textit{Models of Hadron Structure Based on Quantum Chromodynamics}, Springer-Verlag, Berlin (1986); A.J.G. Hey and R.L. Kelly, \textit{Baryon Spectroscopy}, Phys. Rep. \textbf{96}, 71 (1983);   A. Chodos, R.L. Jaffe, K. Johnson and C.B Thorn, \textit{Baryon Structure in the Bag Theory}, Phys. Rev. D \textbf{10}, 2599 (1974).
\bibitem{Shulga} S.G. Shul'ga, and T.P. Il'icheva, \textit{Quasi-potential Approach to the Three-quark Nucleon Wave Function}, Russ. Phys. J.  \textbf{47}, 1242 (2004).
\bibitem{myPRA} M. Donaire, \textit{Net force on an asymmetrically excited two-atom system from vacuum fluctuations}, Phys. Rev. A \textbf{94}, 062701 (2016).


\end{thebibliography}
\end{document}